\documentclass[twocolumn]{article}
\usepackage{geometry}
\usepackage{caption}
\usepackage{graphicx}
\usepackage{amsmath}
\usepackage{amssymb}
\usepackage{textcomp}
\usepackage[dvipsnames]{xcolor}
\usepackage{authblk}
\usepackage{datetime}
\usepackage{gensymb}
\usepackage{wrapfig}
\usepackage{booktabs}
\usepackage{ulem}
\usepackage[numbers]{natbib}
\usepackage{hyperref}
\hypersetup{
    citecolor = blue,
    filecolor = black,
    urlcolor = blue,
    colorlinks = true, 
    linktoc = all,     
    linkcolor = blue,  
}

\newcommand{\onlinecite}[1]{\hspace{-1 ex} \nocite{#1}\citenum{#1}} 
  
\title{Demonstration of Superconducting Optoelectronic\\Single-Photon Synapses}
\author[1]{\large{Saeed Khan$^{\dag1}$, Bryce A. Primavera$^{\dag1,2}$, Jeff Chiles$^1$, Adam N. McCaughan$^1$, Sonia M. Buckley$^1$,\newline Alexander N. Tait$^1$, Adriana Lita$^1$, John Biesecker$^1$, Anna Fox$^1$, David Olaya$^1$,\newline Richard P. Mirin$^1$, Sae Woo Nam$^1$, and Jeffrey M. Shainline$^{*1}$}
\\
\vspace{0.5em}
\textit{\large{$^1$National Institute of Standards and Technology}}
\\
\textit{\large{325 Broadway, Boulder, CO, USA, 80305}}
\vspace{0.5em}
\vspace{0.5em}\\
\textit{\large{$^2$Department of Physics}}\\
\textit{\large{University of Colorado Boulder}}\\
\textit{\large{390 UCB, Boulder, CO, USA, 80309}}\\
\vspace{0.5em}
\small{$^\dag$These authors contributed equally to this work.}\\
\small{$^*$jeffrey.shainline@nist.gov}
}
\date{\today}

\begin{document}

\twocolumn[
\begin{@twocolumnfalse}
\maketitle
\begin{abstract}
Superconducting optoelectronic hardware is being explored as a path towards artificial spiking neural networks with unprecedented scales of complexity and computational ability. Such hardware combines integrated-photonic components for few-photon, light-speed communication with superconducting circuits for fast, energy-efficient computation. Monolithic integration of superconducting and photonic devices is necessary for the scaling of this technology. In the present work, superconducting-nanowire single-photon detectors are monolithically integrated with Josephson junctions for the first time, enabling the realization of superconducting optoelectronic synapses. We present circuits that perform analog weighting and temporal leaky integration of single-photon presynaptic signals. Synaptic weighting is implemented in the electronic domain so that binary, single-photon communication can be maintained. Records of recent synaptic activity are locally stored as current in superconducting loops. Dendritic and neuronal nonlinearities are implemented with a second stage of Josephson circuitry. The hardware presents great design flexibility, with demonstrated synaptic time constants spanning four orders of magnitude (hundreds of nanoseconds to milliseconds). The synapses are responsive to presynaptic spike rates exceeding 10\,MHz and consume approximately 33\,aJ of dynamic power per synapse event before accounting for cooling. In addition to neuromorphic hardware, these circuits introduce new avenues towards realizing large-scale single-photon-detector arrays for diverse imaging, sensing, and quantum communication applications.

\vspace{3em}
\end{abstract}
\end{@twocolumnfalse}
]

\setcounter{tocdepth}{1}
\setcounter{secnumdepth}{4}

\section{\label{sec:introduction}Introduction}
Developing the next generation of artificial intelligence is an interdisciplinary endeavor, building upon advancements in computer science, neuroscience, and hardware. Two general lessons have emerged simultaneously from both engineering pursuits and naturalistic inquiry: (1) a neural system's performance typically improves with increasing number and connectivity of processing primitives, both in biology \cite{dicke2016neuronal, herculano2009human, sterling2015principles} and artificial intelligence \cite{hestness2017deep, brown2020language}; and (2) analog processing with spiking communication in systems exhibiting complex temporal dynamics is both physically efficient and computationally powerful, again in the biological \cite{koch2000role, laughlin2003communication, sterling2015principles} or artificial domain \cite{schemmel2010wafer, indiveri2011neuromorphic}. The circuits presented in this work are designed to embrace both of these ideas by implementing synapses that receive single-photon communication events and perform analog computation with superconducting electronics.

Spiking neural networks inspired by biological systems are particularly intriguing for their exploitation of the temporal domain and suitability for efficient implementation in analog hardware. With recent advances in training algorithms \cite{zenke2018superspike, kaiser2020synaptic, tavanaei2019deep}, spiking networks are increasingly competitive with conventional neural networks and have been argued to be optimal in at least some respects \cite{beer2020spiking, indiveri2019importance}. Implementing sophisticated spike-based processing in large, highly interconnected networks desired for the highest-performance applications remains daunting for current hardware. While biological neurons are capable of directly fanning out to tens of thousands of synapses, present-day electronic systems struggle with physical fan-out greater than a few and typically resort to digital multiplexing \cite{liu2014event}. Multiplexed communication systems inevitably introduce trade-offs between network size and latency. Direct connections between neurons are therefore desirable but require novel hardware. Our proposed solution is to avoid multiplexed communication through the use of integrated optical receivers and transmitters, which do not suffer from charge-based parasitics and can achieve dedicated connections from each neuron to thousands of synaptic recipients. Dense photonic waveguide networks \cite{chiles2017multi,chiles2018design} enable high fan-out neurons with light-speed communication. We have argued elsewhere that superconducting hardware is uniquely promising for realizing large-scale spiking neural networks with sophisticated processing units \cite{shainline2019superconducting,shainline2021optoelectronic,primavera2021considerations}. Superconducting single-photon detectors (SPDs) enable the optical communication of synaptic events at the physical limit of energy efficiency, while the speed, nonlinearity, and low power of Josephson junctions (JJs) make them highly attractive for implementing neural behavior, as recognized by many authors for decades \cite{harada1991artificial,hidaka1991artificial,mizugaki1994implementation,rippert1997multilayered,kondo2005design,hirose2007pulsed,crotty2010josephson,schneider2022supermind}.

SPDs and JJs are combined in this study to realize the first superconducting optoelectronic synapses for large-scale neuromorphic hardware. The synapses are demonstrated to detect single-photon events and implement analog signal weighting in the electronic domain. In addition to these minimum synaptic-processing requirements, the synapses perform leaky integration of events over time, and the non-dissipative nature of superconductivity allows leak rates to be chosen over a wide range of timescales (hundreds of nanoseconds to milliseconds are demonstrated here). This capability is promising for leveraging temporal dynamics over many orders of magnitude in complex, adaptive networks. Synaptic circuits are also shown to inductively couple to neuronal and dendritic circuit blocks for further non-linear processing in a manner that has been shown theoretically to enable high fan-in \cite{primavera2021active}. Taken together and recognizing the synapse as the fundamental computational element of neural systems \cite{shepherd2004synaptic}, these results are a major advancement towards the development of large-scale superconducting optoelectronic networks. To our knowledge, the present work is the first demonstration of monolithic integration of superconducting nanowire single photon detectors with JJs, which will also find application in a variety of other scientific and technological fields.

\section{\label{sec:circuit}The Synaptic Circuit}

\begin{figure*}[t!]
\centering
\includegraphics[width=17.2cm]{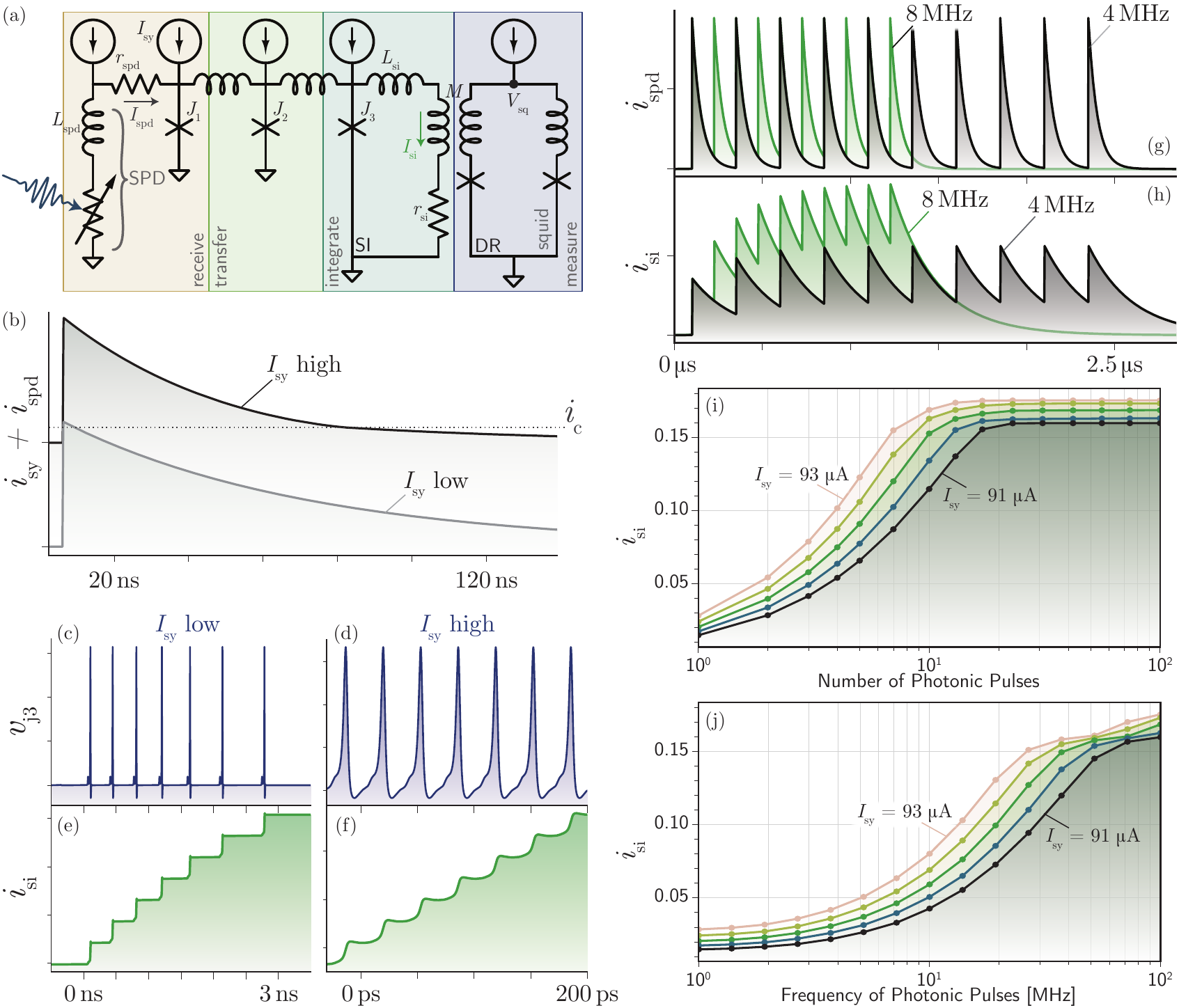}
\caption{Synapse concept. Refer to Appendix\,\ref{apx:circuit_model} for details of the circuit model. (a) Synaptic circuit diagram showing four main circuit blocks. (b) Simulation of the current flowing into $J_1$ upon detection of a photon for two different values of $I_\mathrm{sy}$. A train of fluxons is produced for as long as these traces exceeds the critical current $I_\mathrm{c}$ of the junction. (c)-(f) Fluxon trains for the two synaptic weights and their corresponding additions to the SI loop. The high synaptic weight case, (d) and (f), only shows a brief fraction of the pulse train. (g) SPD response to a train of single-photon pulses at different frequencies. (h) SI current in response to the two photonic pulse trains shown in (g). The higher frequency input pulse train results in a higher current in the integration loop. (i) The peak value of SI current (normalized by the JJ critical current) as function of the number of pulses in a pulse train, demonstrating the emergence of steady-state behavior and synaptic weighting capability. (j) The maximum SI current as a function of the frequency of a pulse train, illustrating the saturating behavior of the SI loop. (i) and (j) use time-accelerated SPDs to reduce numerical simulation time.}
\label{fig:circuit_concept}
\end{figure*}

The basic synaptic circuit is shown in Fig.\,\ref{fig:circuit_concept}(a) and is discussed in detail in Refs.\,\onlinecite{shainline2019superconducting}, \onlinecite{shainline2018circuit}, and \onlinecite{shainline2019fluxonic}. Example behavior shown in Figs.\,\ref{fig:circuit_concept}(b)-(j) was calculated with the circuit model described in Appendix \ref{apx:circuit_model}. The circuit comprises an SPD receiver at the left and a signal integrator at the right. We refer to the integrator as the synaptic integration (SI) loop. Each time the SPD detects one or more photons, current ($I_\mathrm{si}$) is added to the SI loop, and the amount of current added is independent of the number of photons detected, yielding binary photonic communication. This current is added in discrete increments called fluxons \cite{tinkham2004introduction,vatu1998} by the JJ circuit. In the quiescent state, the bias $I_\mathrm{sy}$ is chosen so that the current flowing through the first junction ($J_1$) is below that junction's critical current, $I_\mathrm{c}$, and the junction provides a superconducting path to ground. Upon detection of a photon, the SPD transitions to a resistive state and diverts current $I_\mathrm{spd}$ into $J_1$. When $I_\mathrm{sy} + I_\mathrm{spd}$ exceeds $I_\mathrm{c}$, $J_1$ will produce a train of fluxons that propagates through a Josephson transmission line and accumulates in the SI loop. The synapse makes use of passive reset, as supercurrent will return to the SPD with a time constant set by $L_\mathrm{spd}/r_\mathrm{spd}\approx 40$\,ns. The number of fluxons added to the SI loop per photon-detection event depends on the duration of time $J_1$ is driven above $I_\mathrm{c}$. This behavior is illustrated in Fig.\,\ref{fig:circuit_concept}(b), where $I_\mathrm{sy} + I_\mathrm{spd}$ is plotted for two different values of $I_\mathrm{sy}$. The dashed horizontal line in Fig.\,\ref{fig:circuit_concept}(b) represents the junction $I_\mathrm{c}$. Fluxons are produced for as long as $I_\mathrm{spd} + I_\mathrm{sy}$ remains above $I_\mathrm{c}$. For the low value of $I_\mathrm{sy}$, the combined currents exceed $I_\mathrm{c}$ only briefly, while the high value drives $J_1$ above $I_\mathrm{c}$ for nearly the entire duration of the SPD pulse. Additionally, the rate of fluxon production increases with the current flowing through the junction. Example fluxon trains for the two synaptic weights are shown in Fig.\,\ref{fig:circuit_concept}(c) and Fig.\,\ref{fig:circuit_concept}(d), and their corresponding contributions of current to the SI loop are illustrated in Fig.\,\ref{fig:circuit_concept}(e) and Fig.\,\ref{fig:circuit_concept}(f). Only a brief subsection of the fluxon train is shown in the high weight case. The difference in fluxon rates is evident, as seven fluxons are produced within 3\,ns by the weak synaptic weight, while only 200\,ps is required to produce seven fluxons for the high synaptic weight. The total number of fluxons added to the SI loop in the case of the synapse event with high synaptic weight is 1346.

The role of fluxon trains has a direct analogy with biological synapses, where the strength of synaptic connections is determined by the number of synaptic vesicles containing neurotransmitters that are passed across the synaptic cleft. In our synapses, the strength of synaptic connections is determined by the number of fluxons passed into the SI loop. As with biology, this low-level, discrete picture can often be disregarded in favor of a simpler, essentially analog description of high level synaptic operation.

Figures \ref{fig:circuit_concept}(g)-(j) concern the integration of multiple photon-detection events in the SI loop. Figure \ref{fig:circuit_concept}(g) shows presynaptic spike trains of photon-detection events at two different frequencies. Each photon-detection event adds current to the SI loop, and the integrated current in the SI loop decays with a time constant set by $\tau_\mathrm{si} = L_\mathrm{si}/r_\mathrm{si}$. We will show that the passive elements $L_\mathrm{si}$ and $r_\mathrm{si}$ can be engineered over many orders of magnitude. In this way, the SI loop exhibits leaky-integrator behavior desired of spiking neural computational primitives, shown in Fig.\,\ref{fig:circuit_concept}(h). In Figs.\,\ref{fig:circuit_concept}(g)-(j) $\tau_\mathrm{si}$ is 250\,ns, and for these figures the model used an accelerated-time synapse with 5\,ns SPD recovery to facilitate numerical efficiency. The slower of the two input photonic pulse trains displayed in Fig.\,\ref{fig:circuit_concept}(g) is 4\,MHz ($1/\tau_\mathrm{si}$), while the faster input train is at twice this frequency. Figure \ref{fig:circuit_concept}(h) shows that the higher-frequency series of 10 pulses results in appreciably larger integrated signal, a feature that will be prominent in the measured data shown in Sec.\,\ref{sec:experiment}. The magnitude of current stored in the SI loop is thus a record of recent synaptic activity that can be used in subsequent computations, including in local weight update circuits.

The synapse as a whole can be modeled with a leaky integrator equation of the form
\begin{equation}
\label{eq:leaky_integrator}
\frac{dI_\mathrm{si}}{dt} = I_\mathrm{fq}\,R_\mathrm{fq}(t) - \frac{I_\mathrm{si}}{\tau_\mathrm{si}},
\end{equation}
where $I_\mathrm{fq} = \Phi_0/L_\mathrm{si}$ is the current of a single flux quantum entering the integration loop, and $R_\mathrm{fq}(t)$ is the rate of flux-quantum production. At a fixed frequency of input photonic pulses, current will accumulate in the SI loop. Yet during each inter-spike interval some of this signal will decay exponentially with time constant $\tau_\mathrm{si}$. A quasi-steady state is reached when the signal added with each photonic pulse counters the signal decay between pulses. Here, we use ``quasi-steady state'' to refer to the circumstance where the time-averaged signal between evenly spaced incident photon pulses is constant from one photon pulse to the next. In this case, the time average of $dI_\mathrm{si}/dt = 0$, and Eq.\,\ref{eq:leaky_integrator} informs us that $\bar{I}_\mathrm{si} = \tau_\mathrm{si}\,I_\mathrm{fq}\,\bar{R}_\mathrm{fq}$, with $\bar{I}_\mathrm{si}$ and $\bar{R}_\mathrm{fq}$ indicating time averages. The time-averaged rate at which fluxons are produced depends on the synaptic weight, the rate at which photons are incident, and the value of $I_\mathrm{si}$. The emergence of steady state behavior is illustrated in Fig.\,\ref{fig:circuit_concept}(i) and its frequency and synaptic weight dependence is illustrated in the transfer function of Fig.\,\ref{fig:circuit_concept}(j). This behavior is experimentally validated in Sec.\,\ref{sec:experiment}.

The SI loop is also inductively coupled to a superconducting quantum interference device (SQUID) to implement non-linear transfer functions and transduce $I_\mathrm{si}$ into a measurable voltage. We refer to this SQUID as the dendritic receiving (DR) loop. The DR transfer function in the present study results in a roughly sigmoidal response and can be tuned with the final bias current. The shape of the response is the culmination of at least three factors: the DR transfer function (presented in Appendix\,\ref{apx:jjs_squids}), the saturating behavior of the SI loop [Figs.\,\ref{fig:circuit_concept}(i) and (j)], and the dead time of the SPD at high frequencies ($\approx$20\,MHz). Although we study individual synapses here, the presence of mutual inductors in the fabrication process allows many synapses to couple into a single DR loop with limited cross-talk to enable large-scale fan-in with active dendritic tree structures \cite{primavera2021active}.

While these synapses have many desirable properties for future large-scale systems, the monolithic integration of JJs and SPDs is the engineering achievement that enabled this demonstration and will also stimulate advances in several applications outside of neuromorphic computing. We now discuss the fabrication process before presenting measured data.

\begin{figure*}[t!]
\centering
\includegraphics[width=17.2cm]{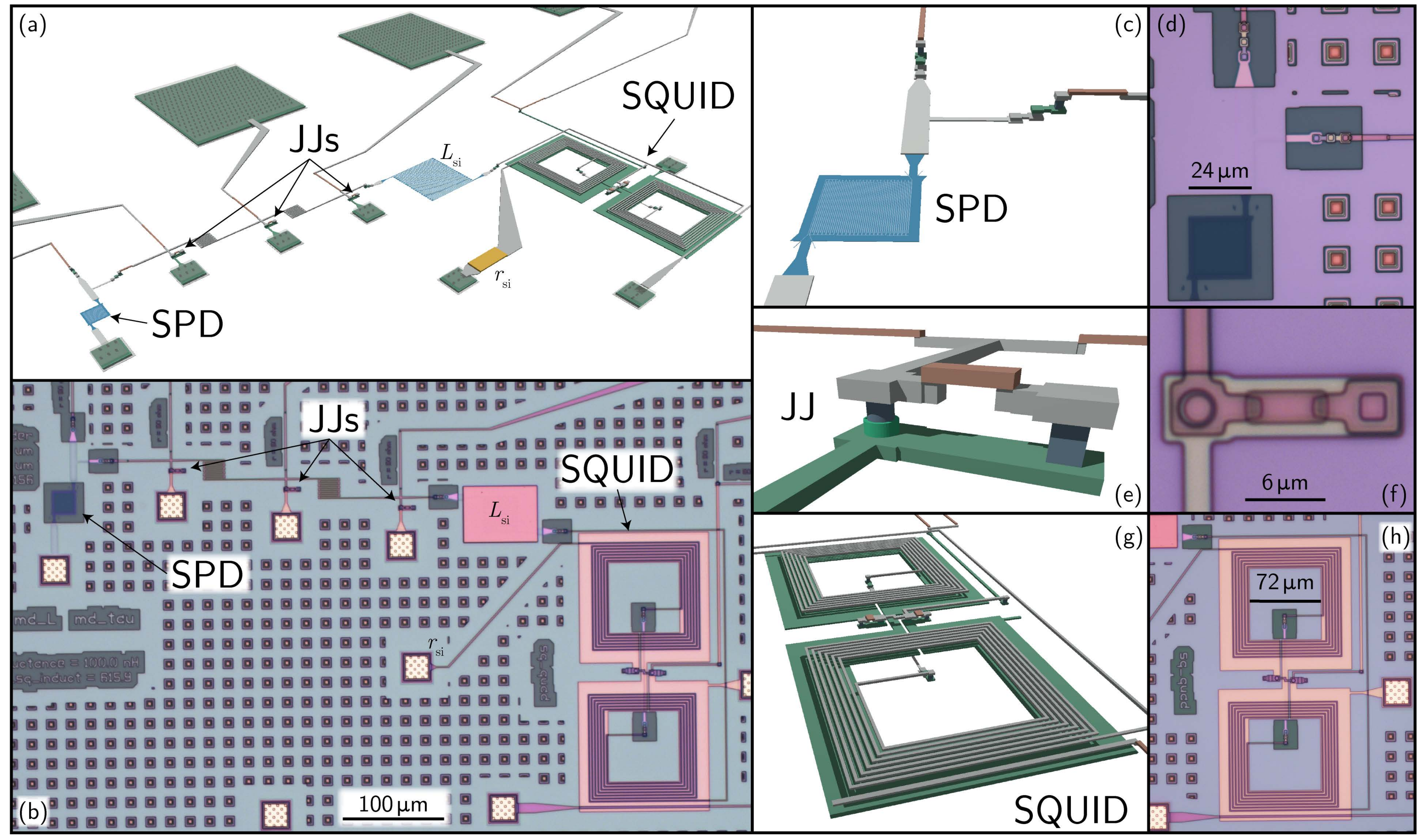}
\caption{Layouts and completed circuits. (a) 3D layout of the entire synapse circuit. (b) Microscope image of completed fabrication. (c)-(h) Device layouts and microscope images for the SPD, JJ, and DR loop.}
\label{fig:layout_uscope}
\end{figure*}

\section{\label{sec:fabrication}Fabrication}

\begin{figure*}[t!]
\centering
\includegraphics[width=17.2cm]{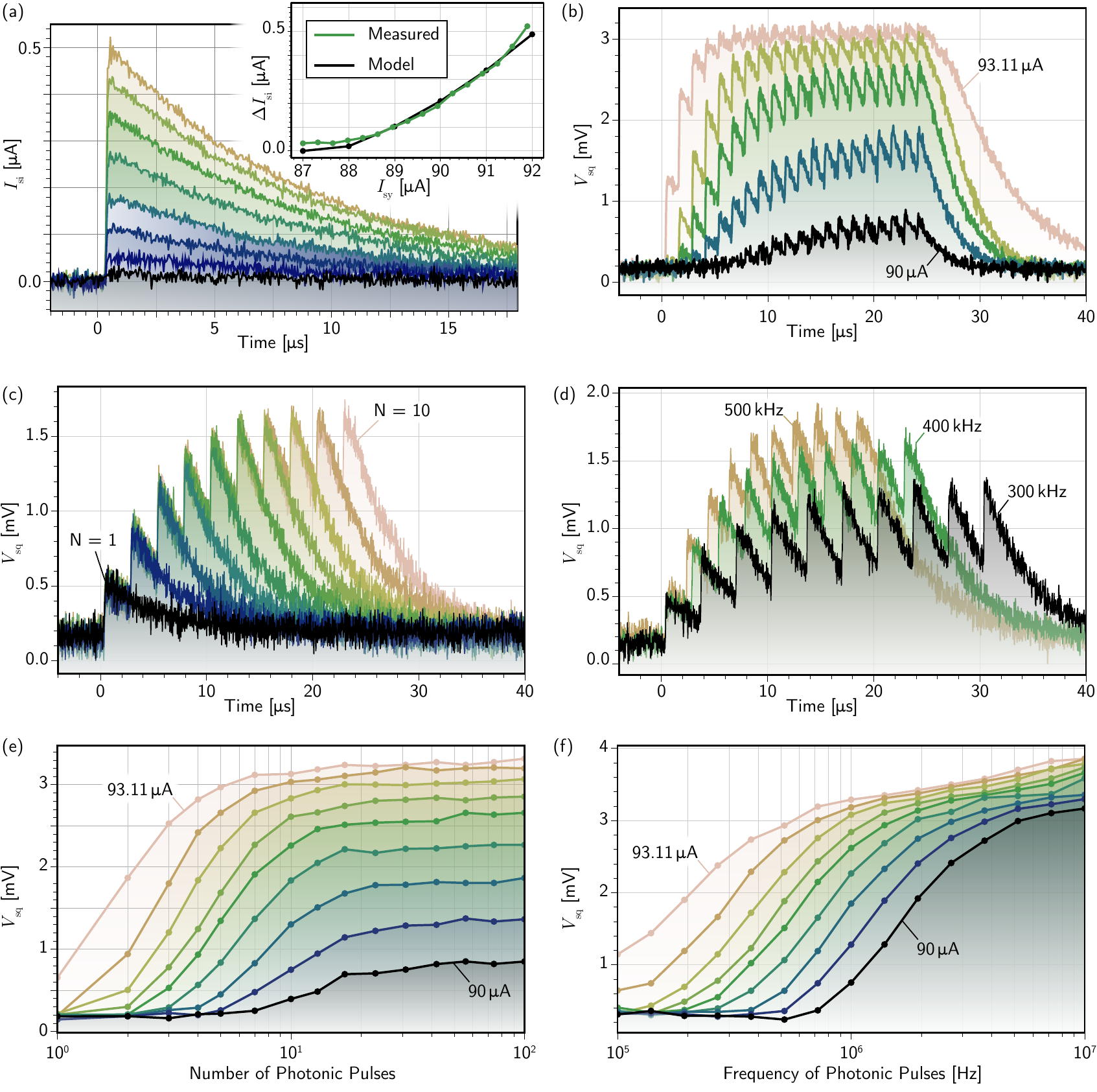}
\caption{Detailed characterization of 6.25\,\textmu s, 2.5\,\textmu H synapse. (a) Response of $I_\mathrm{si}$ to a single optical pulse for different values of $I_\mathrm{sy}$. Inset shows comparison to the model of Appendix\,\ref{apx:circuit_model}. (1000 averages.) (b) Response to an 800\,kHz train of 20 pulses for different synaptic weights. (c) Response at fixed $I_\mathrm{sy}$ to 400\,kHz pulse trains containing different numbers of pulses. (d) Response at fixed $I_\mathrm{sy}$ to pulse trains at three different frequencies. Compare to Fig.\,\ref{fig:circuit_concept}(h). (e) Transfer function versus number of pulses in pulse train at different synaptic weights and fixed frequency (1\,MHz). Compare to Fig.\,\ref{fig:circuit_concept}(i). (f) Transfer function versus pulse train frequency at different synaptic weights for 100 pulses. Compare to Fig.\,\ref{fig:circuit_concept}(j)}
\label{fig:single_synapse_detail}
\end{figure*}

A 14-layer fabrication process was developed for this demonstration and supports Nb/aSi/Nb JJs \cite{olaya2019planarized} externally shunted with PdAu resistors and MoSi SPDs \cite{verma2015high,lita2021mo}. The high kinetic inductance of the MoSi thin film was also leveraged in conjunction with Au resistors to realize the leaky integrating loops. Electron-beam lithography was used for the SPD step, while all other patterning was was accomplished with photolithography using a 365\,nm $i$-line stepper. A complete process flow can be found in Appendix \ref{apx:fabrication_details}. In brief, the MoSi SPD/high-kinetic-inductance layer is deposited and patterned on a pristine oxidized-silicon surface. Contact is made from this layer to Nb. The SPD layer is separated from the JJ layers by an SiO$_2$ insulator and an Nb ground plane. Contact is made from the SPDs to the JJs with etched and backfilled vias. An additional Nb wiring layer is included above the JJs to enable the transformers that couple synapses to SQUIDs. The PdAu and Au resistor layers are patterned last.

A synapse layout and microscope images of key components are shown in Fig.\,\ref{fig:layout_uscope}. Five synapse designs were fabricated with different synaptic time constants and storage capacities. Synapse areas range from 0.32\,mm$^2$-0.52\,mm$^2$ excluding wiring pads, although no effort was made to make the circuits compact in this work. In a mature process, various device layers will be placed atop one another with planarization performed between. Previous analysis found similar synapses will fit in 30\,\textmu m\,$\times$\,30\,\textmu m \cite{primavera2021considerations}. The results presented here are from the first wafer run with this process. The immediate yield is suggestive that the process will be robust.

\section{\label{sec:experiment}Experimental Characterization}
Measurements were performed between 800\,mK and 900\,mK in a closed-cycle sorption pump $^4$He cryostat. Due to wiring limitations, each synapse was measured during a separate cool-down. A fiber-coupled, 780\,nm pulsed laser source was used to flood illuminate the chip and serve as the presynaptic input. The laser pulse width was approximately 480\,ps. This is much shorter than the SPD recovery time, so multiple detection events per pulse are unlikely, as supported by synapse count rate measurements shown in Appendix \ref{apx:spds}. Although each optical pulse contains multiple photons to guarantee detection for each presynaptic event in this free-space configuration, we confirmed the synapses' ability to detect single photons with a linearity measurement under very low light-levels (Appendix \ref{apx:spds}). The detector response is also independent of photon number \cite{buckley2020integrated}, so the fact that more than one photon was used as synaptic input for these measurements is ultimately how the detectors are intended to behave in a network context, and identical dynamics should be expected for few-photon pulses in future waveguide-integrated circuits. As we have argued elsewhere, it may be advantageous for presynaptic spikes to arrive with an average of seven photons to ensure 99\% successful communication despite unavoidable Poisson variability \cite{shainline2021optoelectronic,primavera2021considerations}.

The voltage across the DR loop, $V_\mathrm{sq}$, [Fig.\,\ref{fig:circuit_concept}(a)] is on the order of 10\,\textmu V and is read out with a room-temperature amplifier. Two different amplifiers (40\,dB and 60\,dB voltage gain) were used to accommodate the wide range of timescales under investigation. Averaging was required on most traces to counteract low-frequency line noise disturbing the microvolt signals. An unaveraged trace is shown in Appendix \ref{apx:additional_data}. All biases were generated outside the cryostat and were tuned individually for each synapse to maximize signal amplitude and account for device variation. Refer to Fig.\,\ref{fig:big_table} in Appendix\,\ref{apx:experimental_details} for an exhaustive list of experimental parameters used in the presented data.

Figure\,\ref{fig:single_synapse_detail} presents the characterization of a single synapse designed with a time constant of 6.25\,\textmu s and inductance of 2.5\,\textmu H. Measurements suggest actual values of 8\,\textmu s and 3.2\,\textmu H. The cause of the increased inductance is described in Appendix \ref{apx:fabrication_details}. Synaptic weighting is shown in Fig.\,\ref{fig:single_synapse_detail}(a), where the response to a single optical pulse is plotted for different values of $I_\mathrm{sy}$. The DR response was tuned to be approximately linear over this relatively small signal range, allowing the current in the SI loop, $I_\mathrm{si}$, to be estimated from $V_\mathrm{sq}$ (Appendix\,\ref{apx:jjs_squids}). The inset to Fig.\,\ref{fig:single_synapse_detail}(a) compares the measured data with the circuit model of Appendix \ref{apx:circuit_model} and shows that the weighting function measured experimentally agrees with the prediction of our theoretical model. $I_\mathrm{sy}$ is shown to modulate the height of $V_\mathrm{sq}$ in response to a single optical pulse by at least a factor of 28, and the ability to resolve small weights was limited by noise.

Figure \ref{fig:single_synapse_detail}(b) shows the integrating capability of the synapse in response to a pulse train of fixed frequency and pulse number for five values of synaptic weight. For sufficiently long pulse trains, the synapse reaches a steady state that can be tuned with $I_\mathrm{sy}$, and depends also on the frequency of the input pulse train. We refer to such operation as the ``rate-coding'' domain. Figure \ref{fig:single_synapse_detail}(c) demonstrates an activity level that is better regarded as the ``burst-coding'' domain \cite{zeldenrust2018neural}. Here, the synaptic weight is fixed and the different traces correspond to different numbers of pulses as the steady-state is approached. Figure \ref{fig:single_synapse_detail}(d) shows the synaptic response to 10 pulses at three different frequencies, analogous to the theoretical traces of Fig.\,\ref{fig:circuit_concept}(h). Figures \ref{fig:single_synapse_detail}(e) and \ref{fig:single_synapse_detail}(f) summarize these behaviors by plotting the maximum value of $V_\mathrm{sq}$ as a function of photonic pulse number and frequency respectively for several values of synaptic weight in a manner analogous to the theoretical traces of Fig.\,\ref{fig:circuit_concept}(i) and (j). Figure\,\ref{fig:single_synapse_detail}(e) shows the transition from the burst-coding regime for low numbers of pulses to the rate coding regime where $I_\mathrm{si}$ reaches a steady-state level. Figure\,\ref{fig:single_synapse_detail}(f) displays the desired roll-over behavior with pulse frequency discussed in Sec.\,\ref{sec:circuit}. Both figures also demonstrate the ability of $I_\mathrm{sy}$ to tune synaptic response over a wide range. Sigmoidal fits for these plots are presented in Appendix\,\ref{apx:additional_data}. While the pulse number and frequency transfer functions capture the essential analog computation for the burst- and rate-coding domains, these synapses could be used with spike timing as well \cite{shainline2019fluxonic} and make use of the extremely low jitter (picoseconds) of superconducting nanowire detectors \cite{korzh2020demonstration}.

\begin{figure*}[t!]
\centering
\includegraphics[width=17.2cm]{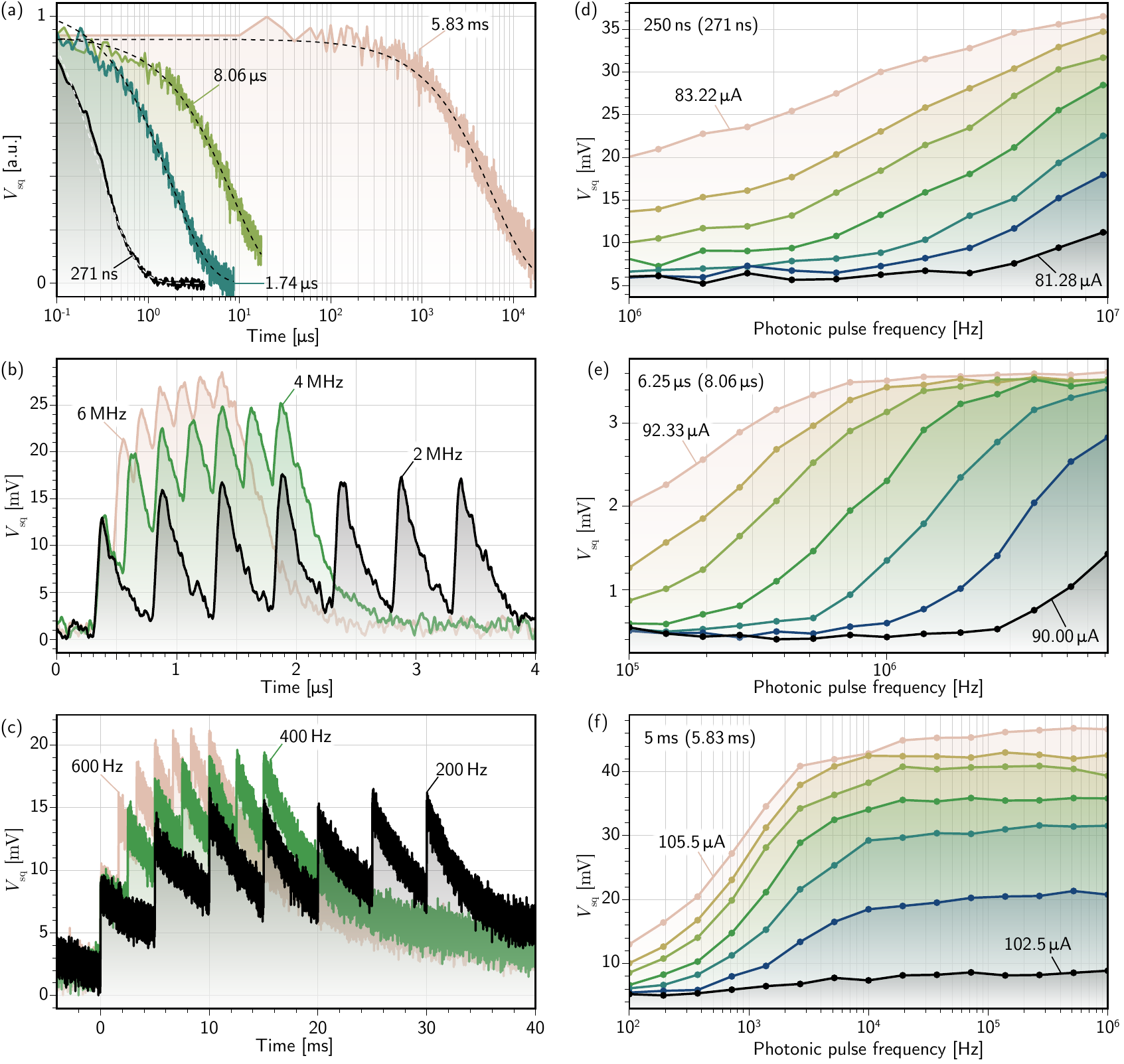}
\caption{Comparison of synapses with varying time constants. (a) Response to a single optical pulse at moderate synaptic weight for synapses with four different designed time constants (250\,ns, 1.25\,\textmu s, 6.25\,\textmu s, and 5\,ms). Dotted lines show exponential fits with extracted time constants. (b) 250\,ns synapse response to seven optical pulses at three different megahertz-range frequencies. (c) 5\,ms synapse response to seven optical pulses at three different sub-kilohertz frequencies. (d)-(f) Frequency transfer function at varying synaptic weights and fixed number of pulses (100, 100, and 10 respectively). Parts (d)-(f) are labeled with the designed time constants, and those measured are given in parentheses.}
\label{fig:time_constant}
\end{figure*}
The synaptic transfer functions can be engineered over a wide range of timescales, as illustrated in Fig.\,\ref{fig:time_constant}. Figure \ref{fig:time_constant}(a) shows the temporal response of synapses with four different time constants to a single photonic pulse. Moderate synaptic weights were chosen for each synapse. We see that the temporal dynamics range from the sub-microsecond to several milliseconds. Dotted lines show exponential fits on the semi-logarithmic plot. Figures \ref{fig:time_constant}(b) and (c) show integrating behavior for the fastest (271\,ns) and slowest (5.83\,ms) time constant synapses respectively, demonstrating no degradation in the leaky-integrating behavior at either extreme. Figures \ref{fig:time_constant}(d)-(f) show the frequency response functions for those same two synapses as well as one with an intermediate time constant designed to be 6.25\,\textmu s and measured to be 8.06\,\textmu s. We once again see similar behavior across the timescales and observe that the onset of integration can be tuned from 100\,Hz to almost 10\,MHz. Figure \ref{fig:time_constant} also illustrates the interplay between parameters fixed in hardware ($L_\mathrm{si}$ and $r_\mathrm{si}$) and those that can be reconfigured dynamically ($I_\mathrm{sy}$).

\begin{figure*}[t!]
\centering
\includegraphics[width=17.2cm]{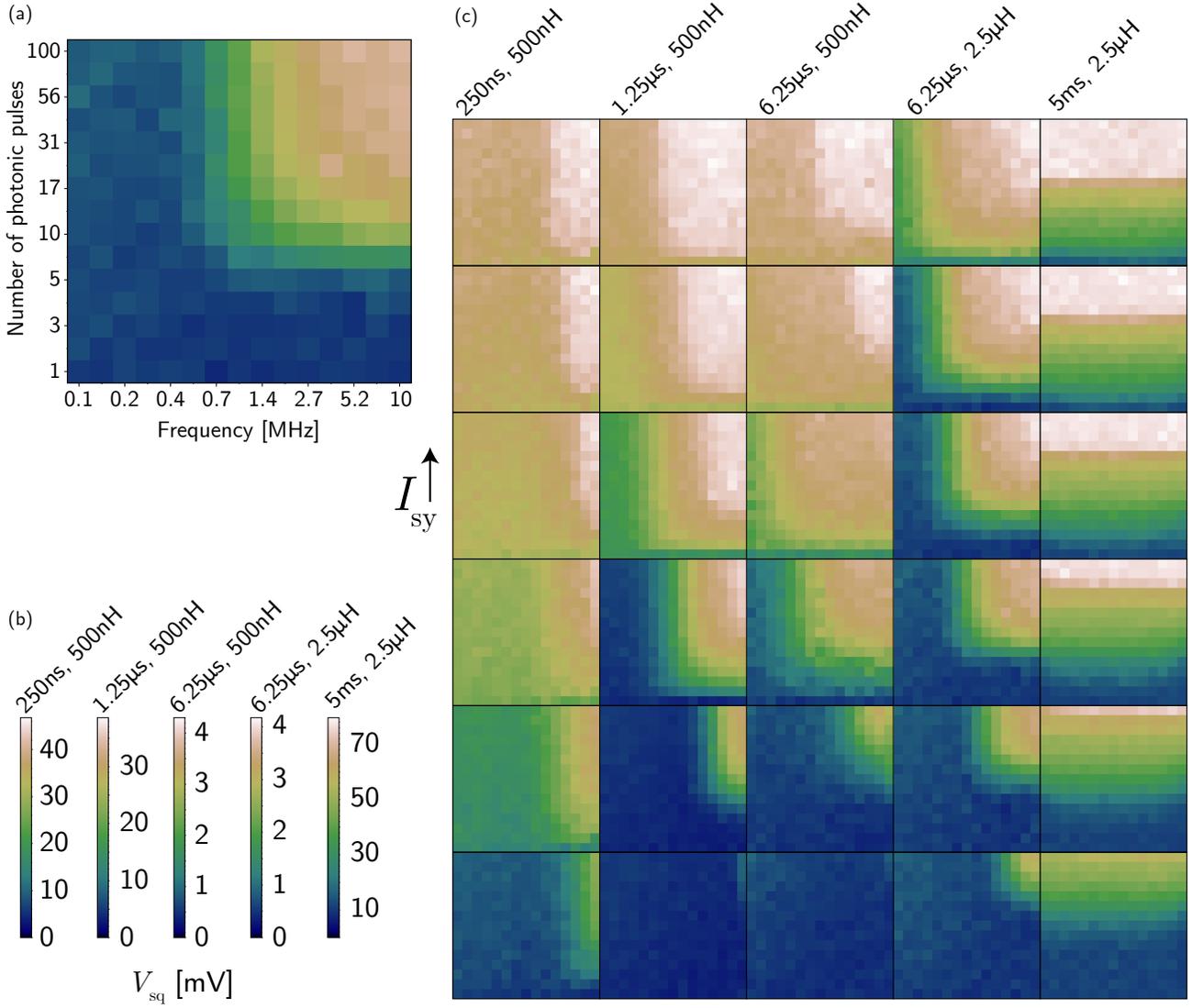}
\caption{Heat maps of synaptic response. (a) Sample map at fixed synaptic weight showing frequency and number axes (fixed for all plots in this figure). (b) Color bars showing the change in $V_\mathrm{sq}$, normalized for each synapse. (c) Grid showing the evolution of synaptic response with $I_\mathrm{sy}$ (increasing vertically) for five different synapses.}
\label{fig:heat_maps}
\end{figure*}
The vast parameter space (pulse frequency, pulse number, $I_\mathrm{sy}$, $\tau_\mathrm{si}$, and $L_\mathrm{si}$) exhibited by these synapses is promising for fostering complex dynamics in large-scale networks. Figure \ref{fig:heat_maps} captures the synaptic response across a large range of this space. The temporal traces of Figures \ref{fig:single_synapse_detail} and \ref{fig:time_constant} are once again reduced to single data points corresponding to the maximum change in voltage recorded for each trace, as in the transfer functions. For fixed $I_\mathrm{sy}$, these values of $V_\mathrm{sq}$ are plotted as 2D heat maps with the frequency of a pulse train along the $x$-axis and the number of pulses in a train along the $y$-axis [Fig. \ref{fig:heat_maps}(a)]. Both $x$ and $y$ axes are spaced with a geometric progression. This data is presented for six different values of $I_\mathrm{sy}$ (rows) and five different synapses (columns) in Fig.\,\ref{fig:heat_maps}(c), and the color bars in Fig.\,\ref{fig:heat_maps}(b) give the scale for each column. Due to the use of different amplifiers and different bias conditions on synapses measured on different cool downs, the variation in color axis was unavoidable. Synapses are labeled by their designed time constant and SI loop inductance. Comparisons between different synapses (columns) should be taken somewhat qualitatively given fabrication and biasing variations. Nonetheless, we can clearly see great diversity in behavior, and several notable trends can be observed. The turn-on frequency increases inversely with $\tau_\mathrm{si}$, as expected for an $LR$ filter. Both the frequency and number responses can be broadly adjusted with synaptic weight. The 250\,ns synapse responds far more strongly to frequencies greater than 1\,MHz, and this is striking in comparison to the microsecond synapses that successfully integrate at much lower frequencies. 

The third and fourth columns show two synapses with the same designed time constant, but two different values of $L_\mathrm{si}$. $L_\mathrm{si}$ sets the amount of current added to the SI loop per fluxon. This is a different weighting mechanism than $I_\mathrm{sy}$, which sets the number of fluxons added per photon detection. Larger $L_\mathrm{si}$ corresponds to less current added to the loop per fluxon, so the apparently slower turn on of the 2.5\,$\mu$H synapse with $I_\mathrm{sy}$ is expected. However, we caution that some of this discrepancy may be from different biasing conditions including a reduced SPD bias on the 2.5\,\textmu H synapse to account for a lower SPD $I_\mathrm{c}$. Investigating the effect of different values of $L_\mathrm{si}$ is intriguing because it sets the ultimate number of events that can be stored by the SI loop, but full analysis requires further study.

\begin{figure*}[t!]
\centering
\includegraphics[width=17.2cm]{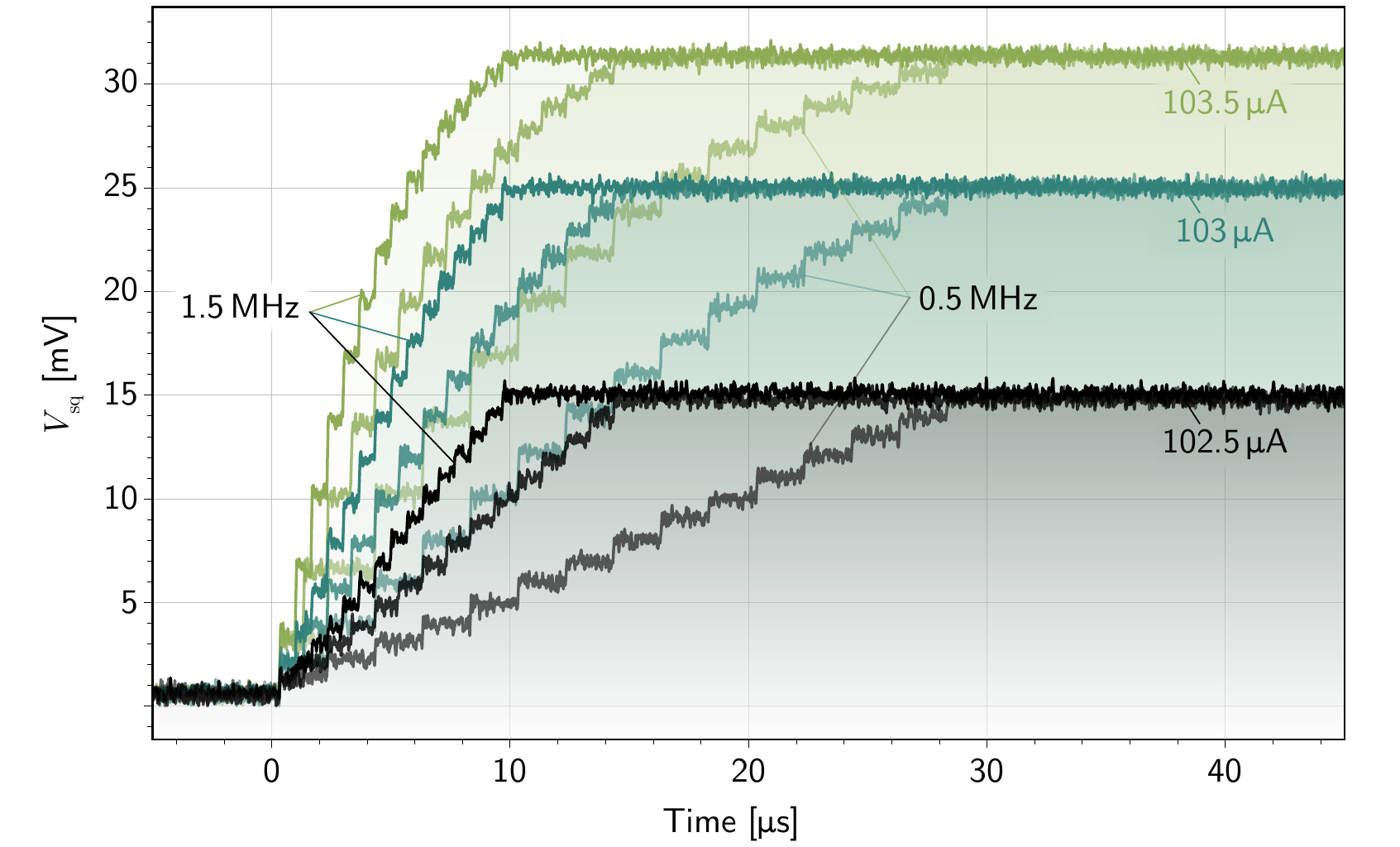}
\caption{The 5\,ms synapse operating in the number-counting regime. There are 15 optical pulses input at 500\,kHz, 1\,MHz, and 1.5\,MHz. We observe that the final value of $V_\mathrm{sq}$ is independent of frequency for this timescale.}
\label{fig:rising_edge}
\end{figure*}
The millisecond synapse in the far right column is a curious case, as it is independent of frequency over this measured range (100\,kHz - 10\,MHz). The current in the SI loop does not have time to decay between pulses, making this synapse essentially a photon-counting device over this range. This slow-leak behavior is illustrated in Fig.\,\ref{fig:rising_edge}, where the millisecond synapse is shown responding to pulse trains of 500\,kHz, 1\,MHz, and 1.5\,MHz. Each pulse train is identical in length (15 events). We see that the final magnitude of $V_\mathrm{sq}$ is independent of frequency, but is instead determined only by $I_\mathrm{sy}$ and the number of pulses in the train. We anticipate this regime to be useful in long-term plasticity mechanisms for modulating the behavior of faster synapses, as well as for applications in SPD imaging arrays.

\section{\label{sec:discussion}Discussion}
Through monolithic integration of SPDs with JJs we have demonstrated tunable single-photon optoelectronic synapses with numerous temporal filtering properties of utility for spiking neural systems. The presented synapses are responsive to presynaptic spiking events encoded as single-photon signals, positioning them for use in large-scale networks that embrace direct optical connections between neurons. Unlike electrical communication, optical signals are immune to the parasitics that pose challenges for electronic fan-out. Direct communication is superior to multiplexed systems for large, highly interconnected networks, as latency is essentially decoupled from network complexity. Latency in this system is limited only by the time it takes light to travel between nodes.

These synapses are not mere variable attenuators as many hardware synapse implementations, but also perform temporal integration and other analog computational primitives observed in biology. A system with integrating synapses (and dendrites) is vastly more dynamic than one in which only the neurons perform integration. The synapses presented here thus make full use of the temporal advantages of spiking networks. The great diversity in demonstrated decay times (hundreds of nanoseconds to several milliseconds) is another advantage of this hardware. We can now imagine networks that are matched to applications of a wide range of different timescales---from accelerated simulations and precise control systems to interactions with humans. The millisecond synapse achieves a biologically relevant timescale, which has been a major area of research in analog CMOS implementations \cite{chicca2014neuromorphic, mayr2015biological}. Mixing synapses with different time constants in the same network may also be advantageous, as networks with time constants spanning orders of magnitude may be more apt to develop temporal dynamics with power-law statistics---the signature of critical behavior that has been studied for its important role in cognition \cite{beggs2008criticality,cocchi2017criticality,tomen2019functional}. Integrating behavior is also useful for plasticity mechanisms that rely on the recent history of synaptic activity to update weights, such as spike-timing-dependent plasticity \cite{kheradpisheh2018stdp, dan2004spike}. We envision local plasticity circuits coupled to the SI loop at every synapse. Many of these plasticity (or homeostatic) mechanisms will also be desired to operate at timescales significantly longer than that of the synapses themselves, which this data suggests is imminently feasible.

This study represents the first demonstration of superconducting optoelectronic synapses, and ample room for improvement remains. The temperature of operation is a prime candidate. The current fabrication process should be commensurate with operating temperatures up to 2.7\,K, but other SPD materials (likely NbTiN) may prove better suited for operation at 4\,K where liquid helium immersion cooling can be utilized. This would not only improve energy efficiency, but would also significantly simplify the measurement apparatus and remove compressor noise. Energy efficiency of the devices themselves also stands to be improved. The main source of energy consumption here is through the discharging of the detector inductance. Free space operation necessitated relatively large detector areas, and thereby large inductance ($\approx$\,825\,nH). This energy ($\frac{1}{2}LI^2$) corresponds to 33\,aJ per event (33\,fJ if a factor of one thousand is included for cooling overhead). A low-inductance waveguide-integrated synapse would improve this metric by an order of magnitude \cite{chiles2017multi}. Optical communication is still likely to dominate power consumption even in this case, so thousands of fluxons can be produced per synaptic operation before computational power begins to dominate over communication. Another major deficiency is the use of an external bias to provide the synaptic weight. A local memory would be superior. One possibility is to store the weight as persistent current in another superconducting loop that is inductively coupled to the $I_\mathrm{sy}$ bias line \cite{shainline2019superconducting}. Such memory should be achievable without any changes to the process flow. Both unsupervised Hebbian-based learning \cite{tavanaei2019deep, diehl2015unsupervised} and recently developed supervised algorithms for local gradient descent \cite{kaiser2020synaptic, bourdoukan2015enforcing} could be implemented with local analog plasticity circuits that adjust the current stored in such memory loops. In the near term, the synapses could be interfaced with bias-generator circuits for either hardware-in-the-loop training or for implementing a fixed network in inference tasks. Further work remains to combine these synapses with light-emitting neuron circuits. Significant progress in cryogenic integrated light sources \cite{buckley2017all, mcdonald2019iii} and superconducting optical transmitter circuits \cite{mccaughan2019superconducting} has occurred, indicating that full superconducting optoelectronic neurons may not be far in the future.

While this platform was designed for high-performance neuromorphic computing, there are many other application areas that stand to benefit from this first demonstration of monolithic SPD-JJ integration. The exact synaptic circuits we have presented can be used as single-photon integrating pixels for advanced imaging applications. Large-scale SPD arrays have thus far been limited by readout technologies, but the integrating character of these circuits allows photon-detection events to be stored and read out at later convenience. The demonstrated millisecond retention times are particularly amenable to large arrays. Single-photon to single-fluxon transduction with long integration times should be feasible, enabling accurate photon counting similar to Ref.\,\onlinecite{onen2019single}. The fabrication process should also be applicable to Josephson circuits incorporating single-flux-quantum logic, presenting the possibility of digital processing of single photon events \cite{yabuno2020scalable, ortlepp2011demonstration} for applications including qubit control or post-processing of images acquired with SPD arrays. Thus, we expect the monolithic SPD-JJ integration to offer new opportunities in fields as diverse as quantum information and communication, biomedical imaging, and broad-spectrum astronomical observations \cite{steinhauer2021progress}, potentially engendering an entirely new field of integrated superconducting optoelectronic hardware.

\section{\label{sec:acknowledgements}Acknowledgements}

\noindent We appreciate recommendations on the readout SQUID design from Dr. Ben Mates, Dr. Malcolm Durkin, and Dr. Jose Aumentado. We appreciate insights on the experimental characterization from Dr. Manuel Castellanos-Beltran and Dana Rampini. This work was made possible by the institutional support of NIST and the effort to advance hardware for artificial intelligence and by the DARPA Invisible Headlights Program.

\bibliographystyle{unsrtnat}
\bibliography{synapse}

\begin{thebibliography}{59}
\providecommand{\natexlab}[1]{#1}
\providecommand{\url}[1]{\texttt{#1}}
\expandafter\ifx\csname urlstyle\endcsname\relax
  \providecommand{\doi}[1]{doi: #1}\else
  \providecommand{\doi}{doi: \begingroup \urlstyle{rm}\Url}\fi

\bibitem[Dicke and Roth(2016)]{dicke2016neuronal}
Ursula Dicke and Gerhard Roth.
\newblock Neuronal factors determining high intelligence.
\newblock \emph{Philosophical Transactions of the Royal Society B: Biological
  Sciences}, 371\penalty0 (1685):\penalty0 20150180, 2016.

\bibitem[Herculano-Houzel(2009)]{herculano2009human}
Suzana Herculano-Houzel.
\newblock The human brain in numbers: a linearly scaled-up primate brain.
\newblock \emph{Frontiers in human neuroscience}, page~31, 2009.

\bibitem[Sterling and Laughlin(2015)]{sterling2015principles}
Peter Sterling and Simon Laughlin.
\newblock \emph{Principles of neural design}.
\newblock MIT press, 2015.

\bibitem[Hestness et~al.(2017)Hestness, Narang, Ardalani, Diamos, Jun,
  Kianinejad, Patwary, Ali, Yang, and Zhou]{hestness2017deep}
Joel Hestness, Sharan Narang, Newsha Ardalani, Gregory Diamos, Heewoo Jun,
  Hassan Kianinejad, Md~Patwary, Mostofa Ali, Yang Yang, and Yanqi Zhou.
\newblock Deep learning scaling is predictable, empirically.
\newblock \emph{arXiv preprint arXiv:1712.00409}, 2017.

\bibitem[Brown et~al.(2020)Brown, Mann, Ryder, Subbiah, Kaplan, Dhariwal,
  Neelakantan, Shyam, Sastry, Askell, et~al.]{brown2020language}
Tom Brown, Benjamin Mann, Nick Ryder, Melanie Subbiah, Jared~D Kaplan, Prafulla
  Dhariwal, Arvind Neelakantan, Pranav Shyam, Girish Sastry, Amanda Askell,
  et~al.
\newblock Language models are few-shot learners.
\newblock \emph{Advances in neural information processing systems},
  33:\penalty0 1877--1901, 2020.

\bibitem[Koch and Segev(2000)]{koch2000role}
Christof Koch and Idan Segev.
\newblock The role of single neurons in information processing.
\newblock \emph{Nature neuroscience}, 3\penalty0 (11):\penalty0 1171--1177,
  2000.

\bibitem[Laughlin and Sejnowski(2003)]{laughlin2003communication}
Simon~B Laughlin and Terrence~J Sejnowski.
\newblock Communication in neuronal networks.
\newblock \emph{Science}, 301\penalty0 (5641):\penalty0 1870--1874, 2003.

\bibitem[Schemmel et~al.(2010)Schemmel, Br{\"u}derle, Gr{\"u}bl, Hock, Meier,
  and Millner]{schemmel2010wafer}
Johannes Schemmel, Daniel Br{\"u}derle, Andreas Gr{\"u}bl, Matthias Hock,
  Karlheinz Meier, and Sebastian Millner.
\newblock A wafer-scale neuromorphic hardware system for large-scale neural
  modeling.
\newblock In \emph{2010 IEEE International Symposium on Circuits and Systems
  (ISCAS)}, pages 1947--1950. IEEE, 2010.

\bibitem[Indiveri et~al.(2011)Indiveri, Linares-Barranco, Hamilton, Schaik,
  Etienne-Cummings, Delbruck, Liu, Dudek, H{\"a}fliger, Renaud,
  et~al.]{indiveri2011neuromorphic}
Giacomo Indiveri, Bernab{\'e} Linares-Barranco, Tara~Julia Hamilton,
  Andr{\'e}~van Schaik, Ralph Etienne-Cummings, Tobi Delbruck, Shih-Chii Liu,
  Piotr Dudek, Philipp H{\"a}fliger, Sylvie Renaud, et~al.
\newblock Neuromorphic silicon neuron circuits.
\newblock \emph{Frontiers in neuroscience}, 5:\penalty0 73, 2011.

\bibitem[Zenke and Ganguli(2018)]{zenke2018superspike}
Friedemann Zenke and Surya Ganguli.
\newblock Superspike: Supervised learning in multilayer spiking neural
  networks.
\newblock \emph{Neural computation}, 30\penalty0 (6):\penalty0 1514--1541,
  2018.

\bibitem[Kaiser et~al.(2020)Kaiser, Mostafa, and Neftci]{kaiser2020synaptic}
Jacques Kaiser, Hesham Mostafa, and Emre Neftci.
\newblock Synaptic plasticity dynamics for deep continuous local learning
  ({DECOLLE}).
\newblock \emph{Frontiers in Neuroscience}, 14:\penalty0 424, 2020.

\bibitem[Tavanaei et~al.(2019)Tavanaei, Ghodrati, Kheradpisheh, Masquelier, and
  Maida]{tavanaei2019deep}
Amirhossein Tavanaei, Masoud Ghodrati, Saeed~Reza Kheradpisheh, Timoth{\'e}e
  Masquelier, and Anthony Maida.
\newblock Deep learning in spiking neural networks.
\newblock \emph{Neural networks}, 111:\penalty0 47--63, 2019.

\bibitem[Beer et~al.(2020)Beer, Urenda, Kosheleva, and
  Kreinovich]{beer2020spiking}
Michael Beer, Julio Urenda, Olga Kosheleva, and Vladik Kreinovich.
\newblock {Why Spiking Neural Networks Are Efficient: A Theorem}.
\newblock In \emph{International Conference on Information Processing and
  Management of Uncertainty in Knowledge-Based Systems}, pages 59--69.
  Springer, 2020.

\bibitem[Indiveri and Sandamirskaya(2019)]{indiveri2019importance}
Giacomo Indiveri and Yulia Sandamirskaya.
\newblock The importance of space and time for signal processing in
  neuromorphic agents: {T}he challenge of developing low-power, autonomous
  agents that interact with the environment.
\newblock \emph{IEEE Signal Processing Magazine}, 36\penalty0 (6):\penalty0
  16--28, 2019.

\bibitem[Liu et~al.(2014)Liu, Delbruck, Indiveri, Whatley, and
  Douglas]{liu2014event}
Shih-Chii Liu, Tobi Delbruck, Giacomo Indiveri, Adrian Whatley, and Rodney
  Douglas.
\newblock \emph{Event-based neuromorphic systems}.
\newblock John Wiley \& Sons, 2014.

\bibitem[Chiles et~al.(2017)Chiles, Buckley, Nader, Nam, Mirin, and
  Shainline]{chiles2017multi}
Jeff Chiles, Sonia Buckley, Nima Nader, Sae~Woo Nam, Richard~P Mirin, and
  Jeffrey~M Shainline.
\newblock Multi-planar amorphous silicon photonics with compact interplanar
  couplers, cross talk mitigation, and low crossing loss.
\newblock \emph{APL Photonics}, 2\penalty0 (11):\penalty0 116101, 2017.

\bibitem[Chiles et~al.(2018)Chiles, Buckley, Nam, Mirin, and
  Shainline]{chiles2018design}
Jeff Chiles, Sonia~M Buckley, Sae~Woo Nam, Richard~P Mirin, and Jeffrey~M
  Shainline.
\newblock Design, fabrication, and metrology of 10\,$\times$\,100 multi-planar
  integrated photonic routing manifolds for neural networks.
\newblock \emph{APL Photonics}, 3\penalty0 (10):\penalty0 106101, 2018.

\bibitem[Shainline et~al.(2019)Shainline, Buckley, McCaughan, Chiles,
  Jafari~Salim, Castellanos-Beltran, Donnelly, Schneider, Mirin, and
  Nam]{shainline2019superconducting}
Jeffrey~M Shainline, Sonia~M Buckley, Adam~N McCaughan, Jeffrey~T Chiles, Amir
  Jafari~Salim, Manuel Castellanos-Beltran, Christine~A Donnelly, Michael~L
  Schneider, Richard~P Mirin, and Sae~Woo Nam.
\newblock Superconducting optoelectronic loop neurons.
\newblock \emph{Journal of Applied Physics}, 126\penalty0 (4):\penalty0 044902,
  2019.

\bibitem[Shainline(2021)]{shainline2021optoelectronic}
Jeffrey~M Shainline.
\newblock Optoelectronic intelligence.
\newblock \emph{Applied Physics Letters}, 118\penalty0 (16):\penalty0 160501,
  2021.

\bibitem[Primavera and
  Shainline(2021{\natexlab{a}})]{primavera2021considerations}
Bryce~A Primavera and Jeffrey~M Shainline.
\newblock Considerations for neuromorphic supercomputing in semiconducting and
  superconducting optoelectronic hardware.
\newblock \emph{Frontiers in Neuroscience}, 15, 2021{\natexlab{a}}.

\bibitem[Harada and Goto(1991)]{harada1991artificial}
Y~Harada and E~Goto.
\newblock Artificial neural network circuits with {J}osephson devices.
\newblock \emph{IEEE transactions on magnetics}, 27\penalty0 (2):\penalty0
  2863--2866, 1991.

\bibitem[Hidaka and Akers(1991)]{hidaka1991artificial}
Mutsuo Hidaka and LA~Akers.
\newblock An artificial neural cell implemented with superconducting circuits.
\newblock \emph{Superconductor Science and Technology}, 4\penalty0
  (11):\penalty0 654, 1991.

\bibitem[Mizugaki et~al.(1994)Mizugaki, Nakajima, Sawada, and
  Yamashita]{mizugaki1994implementation}
Yoshinao Mizugaki, Koji Nakajima, Yasuji Sawada, and Tsutomu Yamashita.
\newblock Implementation of new superconducting neural circuits using coupled
  squids.
\newblock \emph{IEEE transactions on applied superconductivity}, 4\penalty0
  (1):\penalty0 1--8, 1994.

\bibitem[Rippert and Lomatch(1997)]{rippert1997multilayered}
Edward~D Rippert and Susanne Lomatch.
\newblock A multilayered superconducting neural network implementation.
\newblock \emph{IEEE transactions on applied superconductivity}, 7\penalty0
  (2):\penalty0 3442--3445, 1997.

\bibitem[Kondo et~al.(2005)Kondo, Kobori, Onomi, and Nakajima]{kondo2005design}
Taizo Kondo, Masayuki Kobori, Takeshi Onomi, and Koji Nakajima.
\newblock Design and implementation of stochastic neurosystem using {SFQ} logic
  circuits.
\newblock \emph{IEEE transactions on applied superconductivity}, 15\penalty0
  (2):\penalty0 320--323, 2005.

\bibitem[Hirose et~al.(2007)Hirose, Asai, and Amemiya]{hirose2007pulsed}
T~Hirose, T~Asai, and Y~Amemiya.
\newblock Pulsed neural networks consisting of single-flux-quantum spiking
  neurons.
\newblock \emph{Physica C: Superconductivity and its applications},
  463:\penalty0 1072--1075, 2007.

\bibitem[Crotty et~al.(2010)Crotty, Schult, and Segall]{crotty2010josephson}
Patrick Crotty, Dan Schult, and Ken Segall.
\newblock Josephson junction simulation of neurons.
\newblock \emph{Physical Review E}, 82\penalty0 (1):\penalty0 011914, 2010.

\bibitem[Schneider et~al.(2022)Schneider, Toomey, Rowlands, Shainline,
  Tschirhart, and Segall]{schneider2022supermind}
Michael Schneider, Emily Toomey, Graham~E Rowlands, Jeffrey Shainline, Paul
  Tschirhart, and Kenneth Segall.
\newblock Supermind: a survey of the potential of superconducting electronics
  for neuromorphic computing.
\newblock \emph{Superconductor Science and Technology}, 2022.

\bibitem[Primavera and Shainline(2021{\natexlab{b}})]{primavera2021active}
Bryce~A Primavera and Jeffrey~M Shainline.
\newblock An active dendritic tree can mitigate fan-in limitations in
  superconducting neurons.
\newblock \emph{Applied Physics Letters}, 119\penalty0 (24):\penalty0 242601,
  2021{\natexlab{b}}.

\bibitem[Shepherd(2004)]{shepherd2004synaptic}
Gordon~M Shepherd.
\newblock \emph{The synaptic organization of the brain}.
\newblock Oxford university press, 2004.

\bibitem[Shainline et~al.(2018)Shainline, Buckley, McCaughan, Chiles,
  Jafari-Salim, Mirin, and Nam]{shainline2018circuit}
Jeffrey~M Shainline, Sonia~M Buckley, Adam~N McCaughan, Jeff Chiles, Amir
  Jafari-Salim, Richard~P Mirin, and Sae~Woo Nam.
\newblock Circuit designs for superconducting optoelectronic loop neurons.
\newblock \emph{Journal of Applied Physics}, 124\penalty0 (15):\penalty0
  152130, 2018.

\bibitem[Shainline(2019)]{shainline2019fluxonic}
Jeffrey~M Shainline.
\newblock Fluxonic processing of photonic synapse events.
\newblock \emph{IEEE Journal of Selected Topics in Quantum Electronics},
  26\penalty0 (1):\penalty0 1--15, 2019.

\bibitem[Tinkham(2004)]{tinkham2004introduction}
Michael Tinkham.
\newblock \emph{Introduction to superconductivity}.
\newblock Courier Corporation, 2004.

\bibitem[Duzer and Turner(1998)]{vatu1998}
T.~Van Duzer and C.W. Turner.
\newblock \emph{Principles of superconductive devices and circuits}.
\newblock Prentice Hall, USA, second edition, 1998.

\bibitem[Olaya et~al.(2019)Olaya, Castellanos-Beltran, Pulecio, Biesecker,
  Khadem, Lewitt, Hopkins, Dresselhaus, and Benz]{olaya2019planarized}
David Olaya, Manuel Castellanos-Beltran, Javier Pulecio, John Biesecker,
  Soroush Khadem, Theodore Lewitt, Peter Hopkins, Paul Dresselhaus, and Samuel
  Benz.
\newblock {Planarized Process for Single-Flux-Quantum Circuits With
  Self-Shunted Nb/Nb$_{x}$\,Si$_ {1-x}$/Nb Josephson Junctions}.
\newblock \emph{IEEE Transactions on Applied Superconductivity}, 29\penalty0
  (6):\penalty0 1--8, 2019.

\bibitem[Verma et~al.(2015)Verma, Korzh, Bussieres, Horansky, Dyer, Lita,
  Vayshenker, Marsili, Shaw, Zbinden, et~al.]{verma2015high}
Varun~B Verma, Boris Korzh, Felix Bussieres, Robert~D Horansky, Shellee~D Dyer,
  Adriana~E Lita, Igor Vayshenker, Francesco Marsili, Matthew~D Shaw, Hugo
  Zbinden, et~al.
\newblock High-efficiency superconducting nanowire single-photon detectors
  fabricated from {MoSi} thin-films.
\newblock \emph{Optics express}, 23\penalty0 (26):\penalty0 33792--33801, 2015.

\bibitem[Lita et~al.(2021)Lita, Verma, Chiles, Mirin, and Nam]{lita2021mo}
Adriana~E Lita, Varun~B Verma, Jeff Chiles, Richard~P Mirin, and Sae~Woo Nam.
\newblock {Mo$_x$Si$_{1-x}$: A versatile material for nanowire to microwire
  single-photon detectors from UV to near IR}.
\newblock \emph{Superconductor Science and Technology}, 34\penalty0
  (5):\penalty0 054001, 2021.

\bibitem[Buckley et~al.(2020)Buckley, Tait, Chiles, McCaughan, Khan, Mirin,
  Nam, and Shainline]{buckley2020integrated}
Sonia~M Buckley, Alexander~N Tait, Jeffrey Chiles, Adam~N McCaughan, Saeed
  Khan, Richard~P Mirin, Sae~Woo Nam, and Jeffrey~M Shainline.
\newblock Integrated-photonic characterization of single-photon detectors for
  use in neuromorphic synapses.
\newblock \emph{Physical Review Applied}, 14\penalty0 (5):\penalty0 054008,
  2020.

\bibitem[Zeldenrust et~al.(2018)Zeldenrust, Wadman, and
  Englitz]{zeldenrust2018neural}
Fleur Zeldenrust, Wytse~J Wadman, and Bernhard Englitz.
\newblock Neural coding with bursts—current state and future perspectives.
\newblock \emph{Frontiers in computational neuroscience}, 12:\penalty0 48,
  2018.

\bibitem[Korzh et~al.(2020)Korzh, Zhao, Allmaras, Frasca, Autry, Bersin, Beyer,
  Briggs, Bumble, Colangelo, et~al.]{korzh2020demonstration}
Boris Korzh, Qing-Yuan Zhao, Jason~P Allmaras, Simone Frasca, Travis~M Autry,
  Eric~A Bersin, Andrew~D Beyer, Ryan~M Briggs, Bruce Bumble, Marco Colangelo,
  et~al.
\newblock Demonstration of sub-3 ps temporal resolution with a superconducting
  nanowire single-photon detector.
\newblock \emph{Nature Photonics}, 14\penalty0 (4):\penalty0 250--255, 2020.

\bibitem[Chicca et~al.(2014)Chicca, Stefanini, Bartolozzi, and
  Indiveri]{chicca2014neuromorphic}
Elisabetta Chicca, Fabio Stefanini, Chiara Bartolozzi, and Giacomo Indiveri.
\newblock Neuromorphic electronic circuits for building autonomous cognitive
  systems.
\newblock \emph{Proceedings of the IEEE}, 102\penalty0 (9):\penalty0
  1367--1388, 2014.

\bibitem[Mayr et~al.(2015)Mayr, Partzsch, Noack, H{\"a}nzsche, Scholze,
  H{\"o}ppner, Ellguth, and Sch{\"u}ffny]{mayr2015biological}
Christian Mayr, Johannes Partzsch, Marko Noack, Stefan H{\"a}nzsche, Stefan
  Scholze, Sebastian H{\"o}ppner, Georg Ellguth, and Rene Sch{\"u}ffny.
\newblock A biological-realtime neuromorphic system in 28 nm cmos using
  low-leakage switched capacitor circuits.
\newblock \emph{IEEE transactions on biomedical circuits and systems},
  10\penalty0 (1):\penalty0 243--254, 2015.

\bibitem[Beggs(2008)]{beggs2008criticality}
John~M Beggs.
\newblock The criticality hypothesis: how local cortical networks might
  optimize information processing.
\newblock \emph{Philosophical Transactions of the Royal Society A:
  Mathematical, Physical and Engineering Sciences}, 366\penalty0
  (1864):\penalty0 329--343, 2008.

\bibitem[Cocchi et~al.(2017)Cocchi, Gollo, Zalesky, and
  Breakspear]{cocchi2017criticality}
Luca Cocchi, Leonardo~L Gollo, Andrew Zalesky, and Michael Breakspear.
\newblock Criticality in the brain: A synthesis of neurobiology, models and
  cognition.
\newblock \emph{Progress in neurobiology}, 158:\penalty0 132--152, 2017.

\bibitem[Tomen et~al.(2019)Tomen, Herrmann, and Ernst]{tomen2019functional}
Nergis Tomen, J~Michael Herrmann, and Udo Ernst.
\newblock \emph{The functional role of critical dynamics in neural systems},
  volume~11.
\newblock Springer, 2019.

\bibitem[Kheradpisheh et~al.(2018)Kheradpisheh, Ganjtabesh, Thorpe, and
  Masquelier]{kheradpisheh2018stdp}
Saeed~Reza Kheradpisheh, Mohammad Ganjtabesh, Simon~J Thorpe, and Timoth{\'e}e
  Masquelier.
\newblock {STDP}-based spiking deep convolutional neural networks for object
  recognition.
\newblock \emph{Neural Networks}, 99:\penalty0 56--67, 2018.

\bibitem[Dan and Poo(2004)]{dan2004spike}
Yang Dan and Mu-ming Poo.
\newblock Spike timing-dependent plasticity of neural circuits.
\newblock \emph{Neuron}, 44\penalty0 (1):\penalty0 23--30, 2004.

\bibitem[Diehl and Cook(2015)]{diehl2015unsupervised}
Peter~U Diehl and Matthew Cook.
\newblock Unsupervised learning of digit recognition using
  spike-timing-dependent plasticity.
\newblock \emph{Frontiers in computational neuroscience}, 9:\penalty0 99, 2015.

\bibitem[Bourdoukan and Deneve(2015)]{bourdoukan2015enforcing}
Ralph Bourdoukan and Sophie Deneve.
\newblock Enforcing balance allows local supervised learning in spiking
  recurrent networks.
\newblock \emph{Advances in Neural Information Processing Systems}, 28, 2015.

\bibitem[Buckley et~al.(2017)Buckley, Chiles, McCaughan, Moody, Silverman,
  Stevens, Mirin, Nam, and Shainline]{buckley2017all}
Sonia Buckley, Jeffrey Chiles, Adam~N McCaughan, Galan Moody, Kevin~L
  Silverman, Martin~J Stevens, Richard~P Mirin, Sae~Woo Nam, and Jeffrey~M
  Shainline.
\newblock All-silicon light-emitting diodes waveguide-integrated with
  superconducting single-photon detectors.
\newblock \emph{Applied Physics Letters}, 111\penalty0 (14):\penalty0 141101,
  2017.

\bibitem[McDonald et~al.(2019)McDonald, Moody, Nam, Mirin, Shainline,
  McCaughan, Buckley, and Silverman]{mcdonald2019iii}
Corey McDonald, Galan Moody, Sae~Woo Nam, Richard~P Mirin, Jeffrey~M Shainline,
  Adam McCaughan, Sonia Buckley, and Kevin~L Silverman.
\newblock {III-V} photonic integrated circuit with waveguide-coupled
  light-emitting diodes and {WSi} superconducting single-photon detectors.
\newblock \emph{Applied Physics Letters}, 115\penalty0 (8):\penalty0 081105,
  2019.

\bibitem[McCaughan et~al.(2019)McCaughan, Verma, Buckley, Allmaras, Kozorezov,
  Tait, Nam, and Shainline]{mccaughan2019superconducting}
Adam~N McCaughan, Varun~B Verma, Sonia~M Buckley, JP~Allmaras, AG~Kozorezov,
  AN~Tait, SW~Nam, and JM~Shainline.
\newblock A superconducting thermal switch with ultrahigh impedance for
  interfacing superconductors to semiconductors.
\newblock \emph{Nature electronics}, 2\penalty0 (10):\penalty0 451--456, 2019.

\bibitem[Onen et~al.(2019)Onen, Turchetti, Butters, Bionta, Keathley, and
  Berggren]{onen2019single}
Murat Onen, Marco Turchetti, Brenden~A Butters, Mina~R Bionta, Phillip~D
  Keathley, and Karl~K Berggren.
\newblock Single-photon single-flux coupled detectors.
\newblock \emph{Nano Letters}, 20\penalty0 (1):\penalty0 664--668, 2019.

\bibitem[Yabuno et~al.(2020)Yabuno, Miyajima, Miki, and
  Terai]{yabuno2020scalable}
Masahiro Yabuno, Shigeyuki Miyajima, Shigehito Miki, and Hirotaka Terai.
\newblock Scalable implementation of a superconducting nanowire single-photon
  detector array with a superconducting digital signal processor.
\newblock \emph{Optics express}, 28\penalty0 (8):\penalty0 12047--12057, 2020.

\bibitem[Ortlepp et~al.(2011)Ortlepp, Hofherr, Fritzsch, Engert, Ilin, Rall,
  Toepfer, Meyer, and Siegel]{ortlepp2011demonstration}
T~Ortlepp, M~Hofherr, L~Fritzsch, S~Engert, K~Ilin, D~Rall, H~Toepfer, H-G
  Meyer, and M~Siegel.
\newblock Demonstration of digital readout circuit for superconducting nanowire
  single photon detector.
\newblock \emph{Optics Express}, 19\penalty0 (19):\penalty0 18593--18601, 2011.

\bibitem[Steinhauer et~al.(2021)Steinhauer, Gyger, and
  Zwiller]{steinhauer2021progress}
Stephan Steinhauer, Samuel Gyger, and Val Zwiller.
\newblock Progress on large-scale superconducting nanowire single-photon
  detectors.
\newblock \emph{Applied Physics Letters}, 118\penalty0 (10):\penalty0 100501,
  2021.

\bibitem[Clarke and Braginski(2006)]{clarke2006squid}
John Clarke and Alex~I Braginski.
\newblock \emph{The SQUID handbook: Applications of SQUIDs and SQUID systems}.
\newblock John Wiley \& Sons, 2006.

\bibitem[Olaya et~al.(2014)Olaya, Dresselhaus, and Benz]{olaya2014300}
David Olaya, Paul~D Dresselhaus, and Samuel~P Benz.
\newblock {300-GHz Operation of Divider Circuits Using High-$J_{c}$
  Nb/Nb$_x$Si$_{1-x}$/Nb Josephson Junctions}.
\newblock \emph{IEEE Transactions on Applied Superconductivity}, 25\penalty0
  (3):\penalty0 1--5, 2014.

\bibitem[Toomey et~al.(2018)Toomey, Zhao, McCaughan, and
  Berggren]{toomey2018frequency}
Emily Toomey, Qing-Yuan Zhao, Adam~N McCaughan, and Karl~K Berggren.
\newblock Frequency pulling and mixing of relaxation oscillations in
  superconducting nanowires.
\newblock \emph{Physical Review Applied}, 9\penalty0 (6):\penalty0 064021,
  2018.

\end{thebibliography}

\appendix

\section{\label{apx:circuit_model}Circuit Model}
\begin{figure}[b!]
\centering
\includegraphics[width=8.6cm]{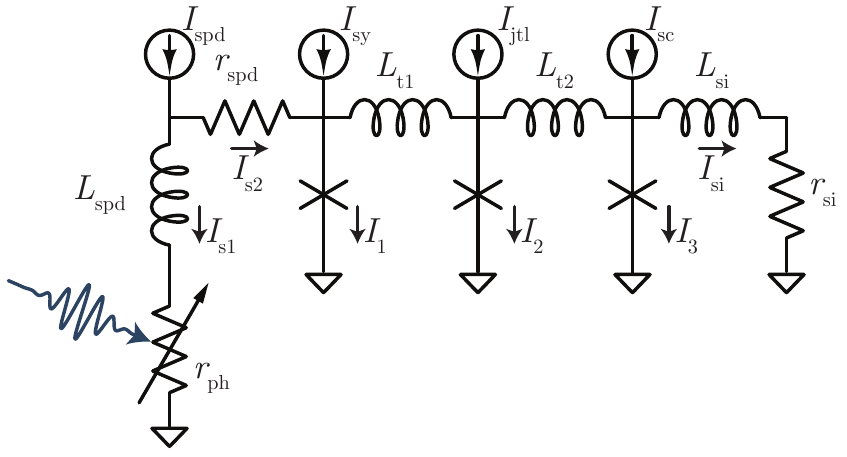}
\caption{Synapse circuit diagram with naming conventions.}
\label{fig:apx_circuit}
\end{figure}
The circuit under consideration is shown in Fig.\,\ref{fig:apx_circuit}. From Josephson's equations, the resistively and capacitively shunted junction model \cite{vatu1998}, and Kirchoff's laws we can derive the following equations of motion for the synapse:
\begin{equation}
\label{eq:odes}
\begin{split}
\frac{d i_\mathrm{s2}}{d\tau} &= \frac{d i_\mathrm{spd}}{d\tau} + \frac{\alpha_{\mathrm{ph}}}{\beta_{\mathrm{Lspd}}} i_{\mathrm{spd}} \\ &-\left( \frac{\alpha_{\mathrm{ph}}+\alpha_{\mathrm{spd}}}{\beta_{\mathrm{Lspd}}} \right) i_{\mathrm{s2}} - \frac{1}{\beta_{\mathrm{Lspd}}} \frac{d \delta_1}{d\tau}, \\
\frac{d i_\mathrm{si}}{d\tau} &= \frac{1}{\beta_{\mathrm{Lsi}}} \frac{d \delta_3}{d\tau} - \frac{\alpha_\mathrm{si}}{\beta_{\mathrm{Lsi}}} i_{\mathrm{si}}.
\end{split}
\end{equation}
Temporal derivatives are with respect to the dimensionless time parameter, $\tau = \omega_c t$, where $\omega_c = 2\pi r_j I_\mathrm{c}/\Phi_0$ is the characteristic frequency of the junction. In Eqs.\,\ref{eq:odes} $\alpha_{\mathrm{x}} = r_{\mathrm{x}}/r_{\mathrm{j}}$ with $x = \mathrm{ph},\;\mathrm{spd}$, or $\mathrm{si}$. $r_{\mathrm{j}}$ is the shunt resistance of the JJs in the resistively and capacitively shunted junction model. The Stewart-McCumber parameter is $\beta_{\mathrm{c}} = 2\pi I_\mathrm{c} r_{\mathrm{j}}^2 c_{\mathrm{j}}/\Phi_0$ with $c_{\mathrm{j}}$ the junction capacitance, and the screening parameter $\beta_{\mathrm{Lx}} = 2\pi I_\mathrm{c} L_x/\Phi_0$ is defined for each inductance (i.e., $x = \mathrm{spd},\;\mathrm{t1},\;\mathrm{t2},\;\mathrm{si}$). Note that this definition of $\beta_{\mathrm{L}}$ differs from that in the SQUID handbook \cite{clarke2006squid} by a factor of $\pi$. In the dimensionless units used here current is normalized to the JJ $I_\mathrm{c}$ ($i_x = I_x/I_\mathrm{c}$). The phase across a JJ is denoted by $\delta$, and each JJ obeys an equation of the form
\begin{equation}
\label{eq:ode_jj}
\frac{d^2 \delta}{d \tau^2} = \frac{1}{\beta_{\mathrm{c}}} \left( i - \mathrm{sin}\delta - \frac{d \delta}{d\tau} \right).
\end{equation}
There are three JJs in the circuit, each obeying the second-order Eq.\,\ref{eq:ode_jj}, and two additional first-order equations from Eqs.\,\ref{eq:odes}. This system can be represented as eight coupled first-order ODEs which we solve with the Scipy function {\fontfamily{cmtt}\selectfont solve\_ivp}. The other currents in the circuit can be obtained from
\begin{equation}
\label{eq:currents}
\begin{split}
i_1 &= \frac{1}{\beta_{\mathrm{Lt1}}} \left( \delta_2 - \delta_1 \right) + i_{\mathrm{sy}} + i_{\mathrm{s2}}, \\
i_3 &= \frac{1}{\beta_{\mathrm{Lt2}}} \left( \delta_2 - \delta_3 \right) - i_{\mathrm{si}} + i_{\mathrm{sc}}, \\
i_2 & = i_{\mathrm{sy}} + i_{\mathrm{jtl}} + i_{\mathrm{sc}} + i_{s2}- i_1 - i_3 - i_{\mathrm{si}}.
\end{split}
\end{equation}

\section{\label{apx:fabrication_details}Fabrication Details}
The fabrication of the SPD-JJ synaptic circuits combines previously developed SPD and JJ fabrication processes. The MoSi SPDs have been described in Refs.\,\onlinecite{verma2015high,lita2021mo}, and the Nb JJs have been described in Refs.\,\onlinecite{olaya2014300, olaya2019planarized}. The process began with a thermally oxidized silicon wafer. Thin Nb contacts (45\,nm) ware deposited via sputtering, and a liftoff process was employed to realize a very shallow sidewall angle. This defined the first metal layer, M1. Next MoSi was sputtered to a thickness of 5\,nm and capped with a 2\,nm amorphous silicon layer. This defined the superconducting thin film layer, STF. The Nb liftoff process was used for M1 so the thin MoSi layer could make superconducting contact. In a mature fabrication it will be desirable to deposit Nb wires with a damascene process followed by chemical-mechanical planarization. The MoSi would be deposited on the planarized surface and would make contact to the planarized Nb. In the present work we avoided this planarization step as it is non-trivial in the NIST cleanroom. We have performed experiments with MoSi deposited on planarized oxidized wafers with less than 0.5\,nm root-mean-square roughness and measured less than 0.2\,K change in the film critical temperature as compared to a film deposited on a thermally oxidized substrate.

The SPDs were defined on the STF layer through ebeam lithography. The wires defining the SPDs were 200\,nm wide on the mask with 50\% fill factor, resulting in wire widths near 180\,nm on the wafer. A 100\,kV column and 2\,nA beam current were used in patterning. The inductors comprising the SI loops were formed from the same STF layer but were defined with $i$-line photolithography. Feature sizes on this layer were 1\,\textmu m or larger. All features on STF were dry etched with SF$_6$ chemistry.

Following M1 and STF, the first SiO$_2$ insulator (I1) was deposited using electron cyclotron resonance plasma-enhanced chemical vapor deposition. I1 was 200\,nm thick, as were all insulating layers. The first via layer (V1) was etched through I1, terminating on the Nb contacts formed in M1. V1 was etched with CHF$_3$ chemistry with O$_2$ added to increase the sidewall slope to facilitate via formation without a damascene process. M2 was then deposited by sputtering Nb, and wires were formed with $i$-line photolithography and dry etching using an inductively coupled plasma with SF$_6$ chemistry. M2 serves as the superconducting ground plane for the circuits. I2 was then deposited using the same process as I1. V2 was opened using the same process as V1.

The JJs were fabricated next using an Nb/a-Si/Nb trilayer. The lower electrode Nb layer (JJ1) made contact with M2 through V2. The a-Si tunneling barrier was approximately 5\,nm thick. Self-shunted JJs with Nb doped Si barriers have been previously used in SFQ circuits \cite{olaya2019planarized}. However, in this work undoped a-Si barriers and external shunts were used as JJ area was not a concern in these circuits. The target critical current density of the trilayer was 1\,kA/cm$^2$, but the resulting value was more than twice as large due to an unidentified change in the process. Subsequent wafers have returned to the previous value. The fact that the synaptic circuits remained highly functional when the critical currents of all JJs on the wafer were more than twice as large as designed is evidence for the robustness of the circuit concepts. The JJ top electrode (JJ2) and tunneling barrier were dry etched with SF$_6$ chemistry, stopping on JJ1 to leave a metal wiring layer. The JJ1 layer was also used as the SQUID washer body, which served as the pickup component of the transformer between the SI and DR loops. I3 was then deposited at 200\,nm with the same SiO$_2$ as the other insulators, and V3 was etched to contact the JJ top and bottom electrodes. 

Upper Nb metal M3 was then deposited and etched just as M2. This wiring layer was used for superconducting interconnects and to form the input coil for the transformer from the SI loop to the DR loop. M1 was also employed in this transformer to cross under JJ1. All SQUIDs on this wafer leveraged quadrupole designs [Fig.\,\ref{fig:layout_uscope}(g) and (h)] to avoid sensitivity to stray uniform magnetic fields. Some SQUIDs on a diagnostic chip included resistors coupled to the washer to damp $LC$ resonances. However, for the SQUIDs employed to interface with the synapses, an abundance of caution drove us to omit those resistors to our detriment. $LC$ resonances driven by the room-temperature amplifier had deleterious effects on the SQUID response curves (Appendix \ref{apx:jjs_squids}), limiting the dynamic range we were able to use for the measurements.

The JJ shunt resistors were formed from PdAu due to that material's low residual resistivity ratio, while the resistors used in the SI loops to set the leak rate were formed from Au due to its low resistivity. These two layers were deposited in separate liftoff steps. Again, a mature foundry would be able to straightforwardly adapt these steps to a damascene process. A final insulator was deposited just as the others, and large vias were opened to enable wire bonding to the top Au layer.

In addition to the JJ tunneling barrier being thinner than desired leading to higher $I_\mathrm{c}$ than designed, one other processing step was sub-optimal. Before depositing I1 over STF, an RF clean was conducted in the deposition chamber. This plasma clean slightly thinned the STF film resulting in suppressed critical temperature and increased inductance. All circuits were designed to operate at 2.7\,K, with the MoSi film having a critical temperature above 5.5\,K, but this film degradation required operation below 1\,K to support the designed SPD $I_\mathrm{c}$ values. The increased inductance is likely the cause of the SI loop inductances being larger than design. This issue has since been resolved by depositing a slightly thicker a-Si capping layer on STF.

All data presented in this paper were acquired from the first wafer fabricated with this combined SPD-JJ process.

\section{\label{apx:experimental_details}Experimental Details}
\begin{figure*}[t!]
\centering
\includegraphics[width=\textwidth]{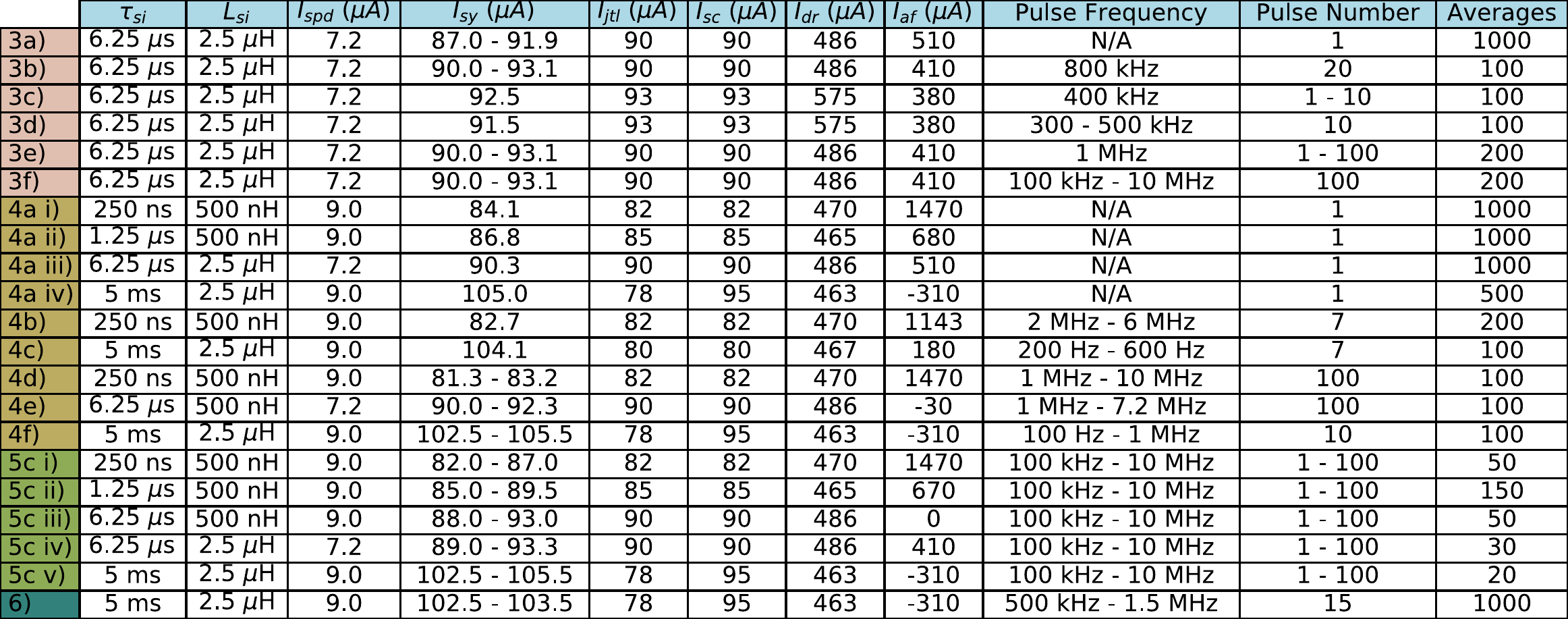}
\caption{Parameters used for all data shown in the main text.}
\label{fig:big_table}
\end{figure*}
\begin{figure*}[t!]
\centering
\includegraphics[width=17.2cm]{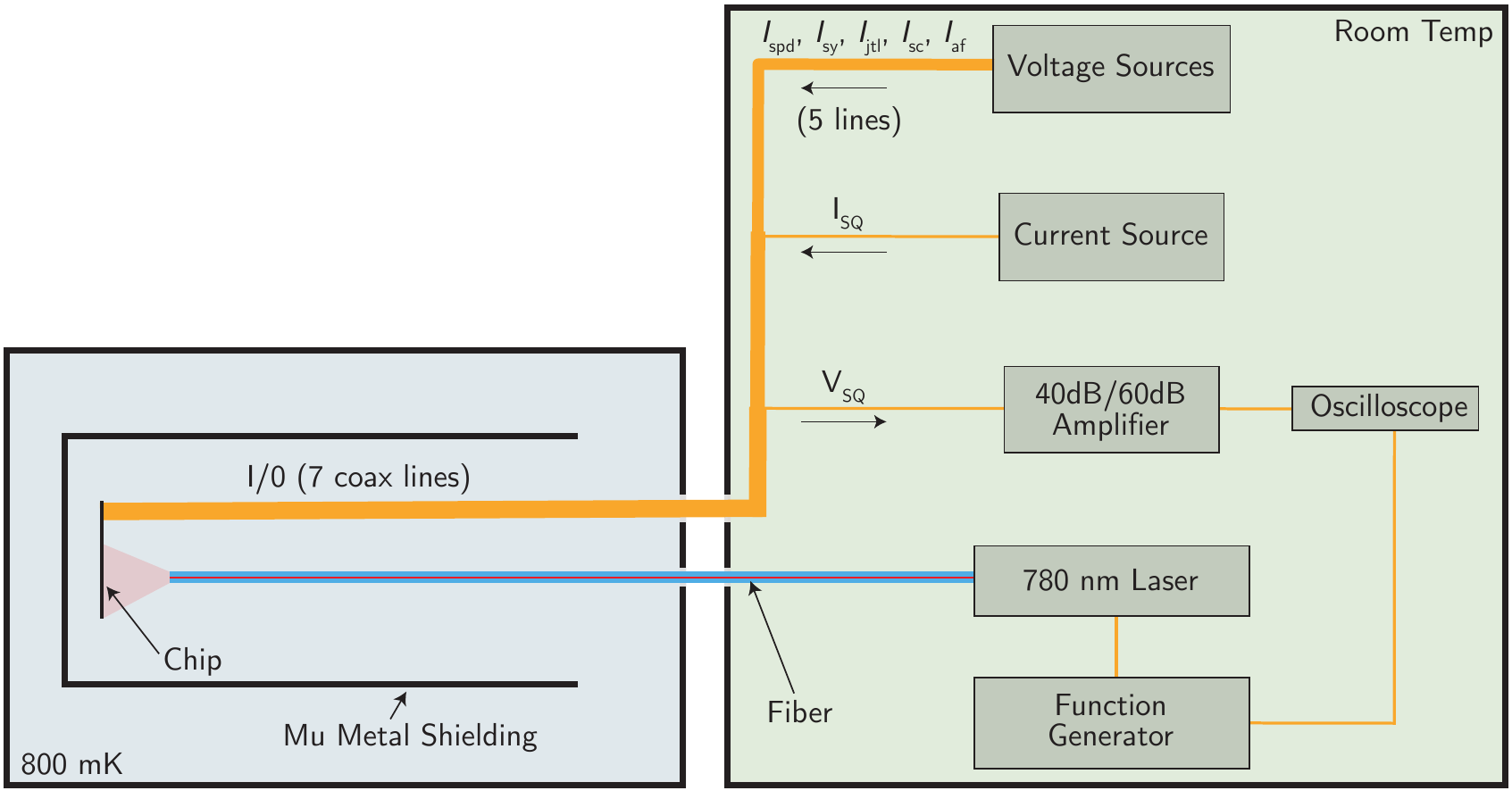}
\caption{Measurement setup. The voltage sources are used to generate currents through fixed resistors between 100\,$\Omega$ and 100\,k$\Omega$.}
\label{fig:set_up}
\end{figure*}

A schematic of the experimental set-up is presented in Fig.\,\ref{fig:set_up}. Each synapse requires seven coaxial connections for I/O. Four current biases are required by the synaptic block ($I_\mathrm{spd}$, $I_\mathrm{sy}$, $I_\mathrm{jtl}$, and $I_\mathrm{sc}$) as shown in Fig.\,\ref{fig:apx_circuit}. The DR block requires a bias for the SQUID ($I_\mathrm{dr}$), an ``addflux'' bias ($I_\mathrm{af}$) for tuning the operating point of the DR loop, and a line for reading the voltage ($V_\mathrm{sq}$) across the loop. Fabrication variation between synapses required each of the six current biases to be adjusted to maximize signal amplitude. Due to current sharing on the JJs between the biases, there are many possible biasing conditions that behave in a qualitatively similar manner. To maintain consistency, the biases $I_\mathrm{sy}$, $I_\mathrm{jtl}$, and $I_\mathrm{sc}$ were chosen with similar values when possible. The addflux line was chosen so that $V_\mathrm{sq}$ just started to increase with $I_\mathrm{si}$, except for where more linear operation was desired, such as in Figs.\,\ref{fig:single_synapse_detail}(a) and \ref{fig:time_constant}(a). The biases for every figure in the main text are provided in Fig.\,\ref{fig:big_table}.

All measurements were performed between 800\,mK and 900\,mK in a closed-cycle sorption pump $^4$He cryostat. Two concentric cylindrical mu-metal shields reduce external magnetic noise. An optical fiber was positioned to flood illuminate the entire chip. A function generator triggered the 780\,nm laser to produce bursts of pulses of a given number and frequency. $I_\mathrm{dr}$ was supplied by a commercially available current source, while all other current biases came from resistors in series with isolated voltage sources. Two different amplifiers were used in these experiments to read $V_\mathrm{sq}$. One was a home-built amplifier with 40\,dB voltage gain and 1\,MHz bandwidth. The other was a commercial amplifier with 60\,dB gain and 10\,MHz bandwidth. Amplifier bandwidth limitations are visible for the 250\,ns synapse [Fig.\,\ref{fig:time_constant}(b)]. Time traces were recorded with a 1\,GHz oscilloscope triggered by the function generator.

\section{\label{apx:spds}Characterization of Single-Photon Detectors}
\begin{figure}[t!]
\centering
\includegraphics[width=8.6cm]{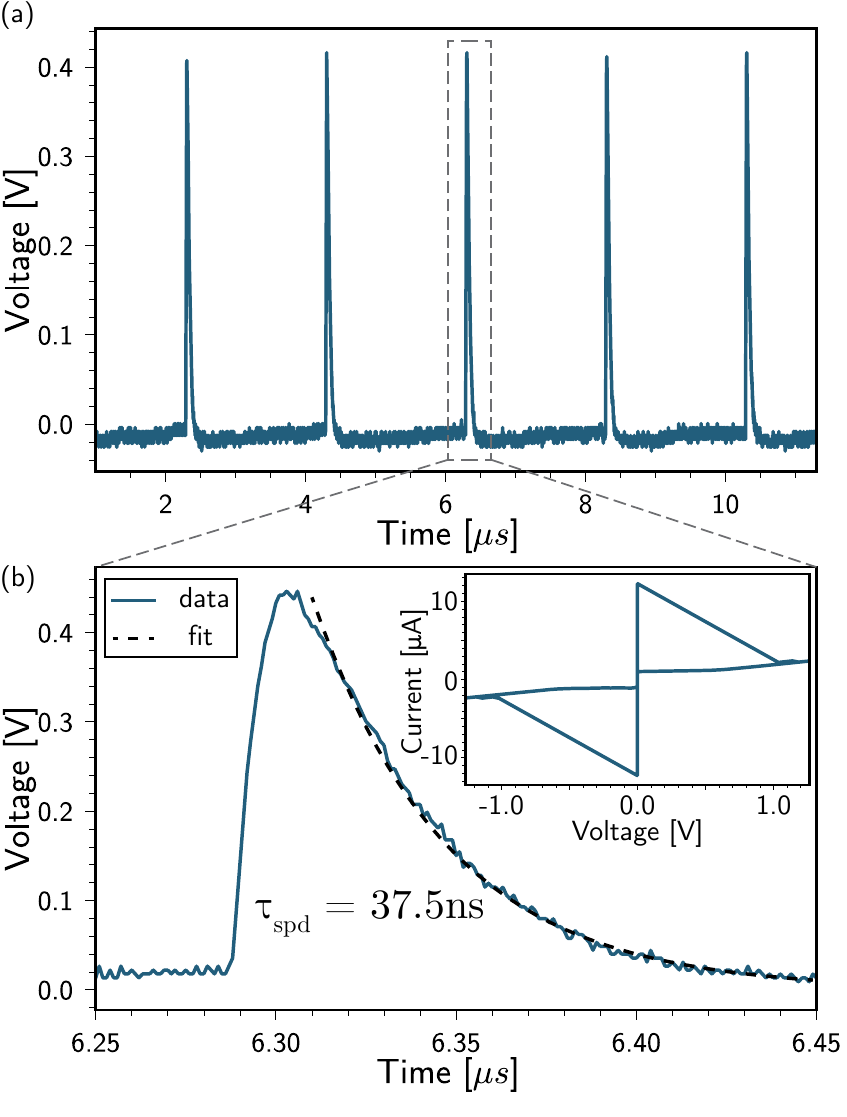}
\caption{Monitor SPD counts. (a) Response to a 500\,kHz pulse train. (b) Response to a single pulse with exponential fit. Inset shows monitor SPD IV curve.}
\label{fig:spd_diagnostics_1}
\end{figure}
\begin{figure}[t!]
\centering
\includegraphics[width=8.6cm]{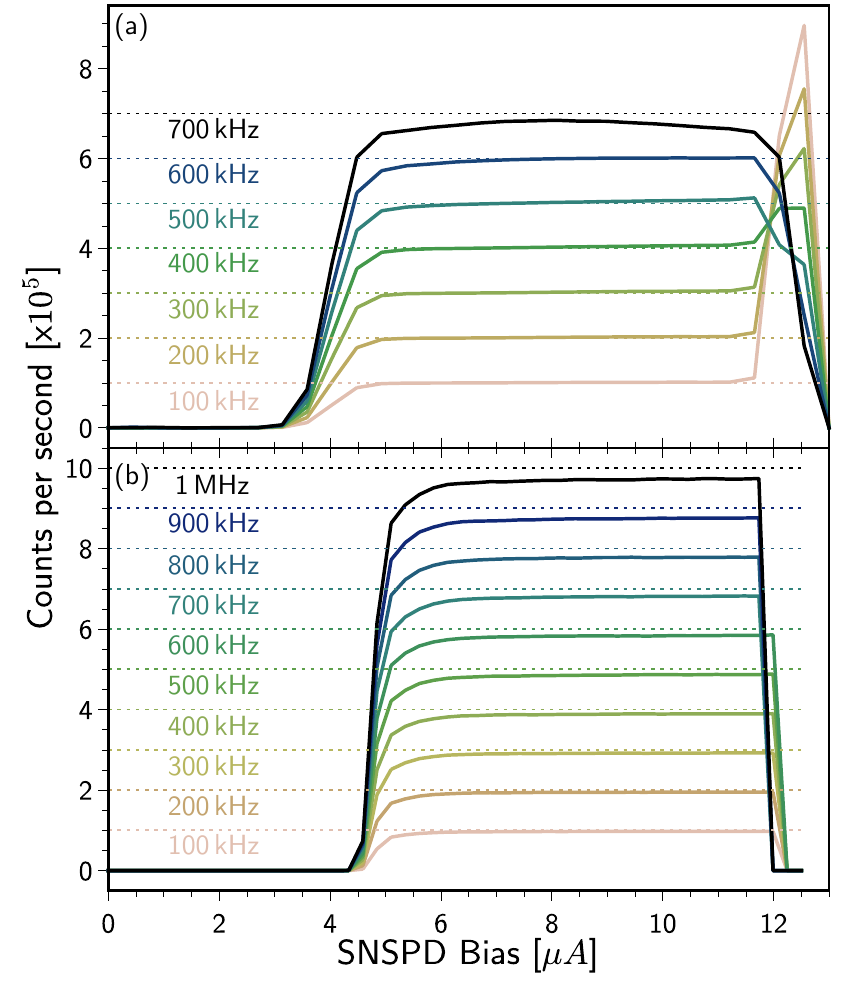}
\caption{Count rate versus SPD bias current for different frequencies of input pulses for (a) the 250\,ns synapse and (b) the monitor SPD.}
\label{fig:spd_diagnostics_3}
\end{figure}

The SPDs are MoSi superconducting-nanowire single-photon detectors \cite{lita2021mo}. A typical detector response to a photon pulse train is shown in Fig.\,\ref{fig:spd_diagnostics_1}(a) and a zoom-in of a single pulse is shown in Fig.\,\ref{fig:spd_diagnostics_1}(b). The decay time is approximately 37.5\,ns. The SNSPD is unresponsive to photons when in the resistive or recovery state, so this response time sets the maximum frequency (tens of MHz) of presynaptic events for this hardware. The inset of Fig.\,\ref{fig:spd_diagnostics_1}(b) shows the IV curve of the detector, demonstrating a critical current of approximately 12\,\textmu A. Figure \ref{fig:spd_diagnostics_3} shows count rate as a function of bias current for pulse trains of various frequencies for both the 250\,ns synapse (a) and a stand-alone ``monitor'' SPD (b). Such a monitor SPD was fabricated adjacent to each synapse for diagnostic purposes. The pulse frequency is kept below the onset of integration for the synapse so that individual counts could be well-defined. We observe a large plateau region from 6\,\textmu A-12\,\textmu A where the photon detection efficiency is largely independent of bias (and bias noise). The synapse exhibits a large increase in dark counts near $I_\mathrm{c}$ likely due to relaxation oscillations between the SPD and the Josephson junctions \cite{toomey2018frequency}. Fortunately, this phenomenon only occurs for a small range of biases, and is easily avoidable by biasing elsewhere in the plateau. SPD biases in this work were 9\,\textmu A for all synapses except the 6.25\,\textmu s, 2.5\,\textmu H synapse where 7.2\,\textmu A was used to account for a reduced SPD critical current.

\begin{figure}[t!]
\centering
\includegraphics[width=8.6cm]{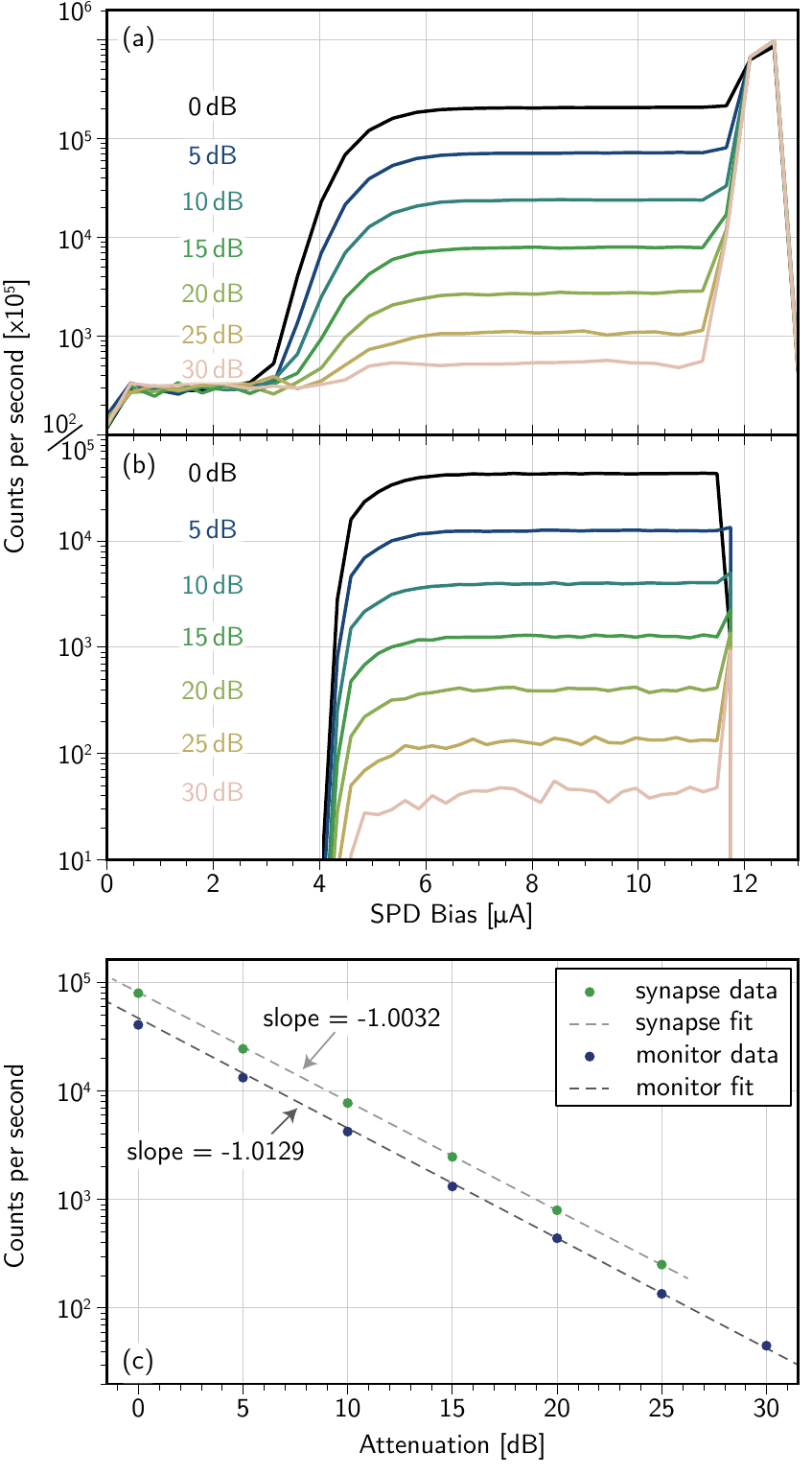}
\caption{Count rate dependence on optical power level. (a) Count rate for the 250\,ns synapse as a function of SPD bias current. (b) Count rate for the monitor SPD as a function of bias. (c) Confirmation of single-photon sensitivity for both the monitor SPD and the synapse.}
\label{fig:spd_diagnostics_2}
\end{figure}

The single-photon operation of the detectors and synapses is confirmed in Fig.\,\ref{fig:spd_diagnostics_2}(c). Under very low optical illumination, single photon events are expected to dominate optical inputs, as opposed to multi-photon events wherein more than one photon is absorbed in the detector within the $\approx 200$\,ps rise time of the SPD pulse. For optical pulses with $\lambda$ photons per pulse, the probability of detection is proportional to the Poisson probability of a single photon in a pulse, $P(k = 1) = \lambda e^{-\lambda}$. For $\lambda \ll 1$, Taylor expansion shows that the count rate should be proportional to $\lambda$, and therefore to the optical power. Taking the logarithm of both sides implies that the logarithm of count rate versus the logarithm of power should be a line with a slope of positive one. This is confirmed in Fig.\,\ref{fig:spd_diagnostics_2}(c), where linear fits to the logarithm of count rate versus attenuation show slopes very near the expected value of negative one for both the synapse and monitor SPD. If two-photon events were required for detection, the slope would be negative two.

\section{\label{apx:jjs_squids}Characterization of Josephson Junctions and Superconducting Quantum Interference Devices}
\begin{figure}[t!]
\centering
\includegraphics[width=8.6cm]{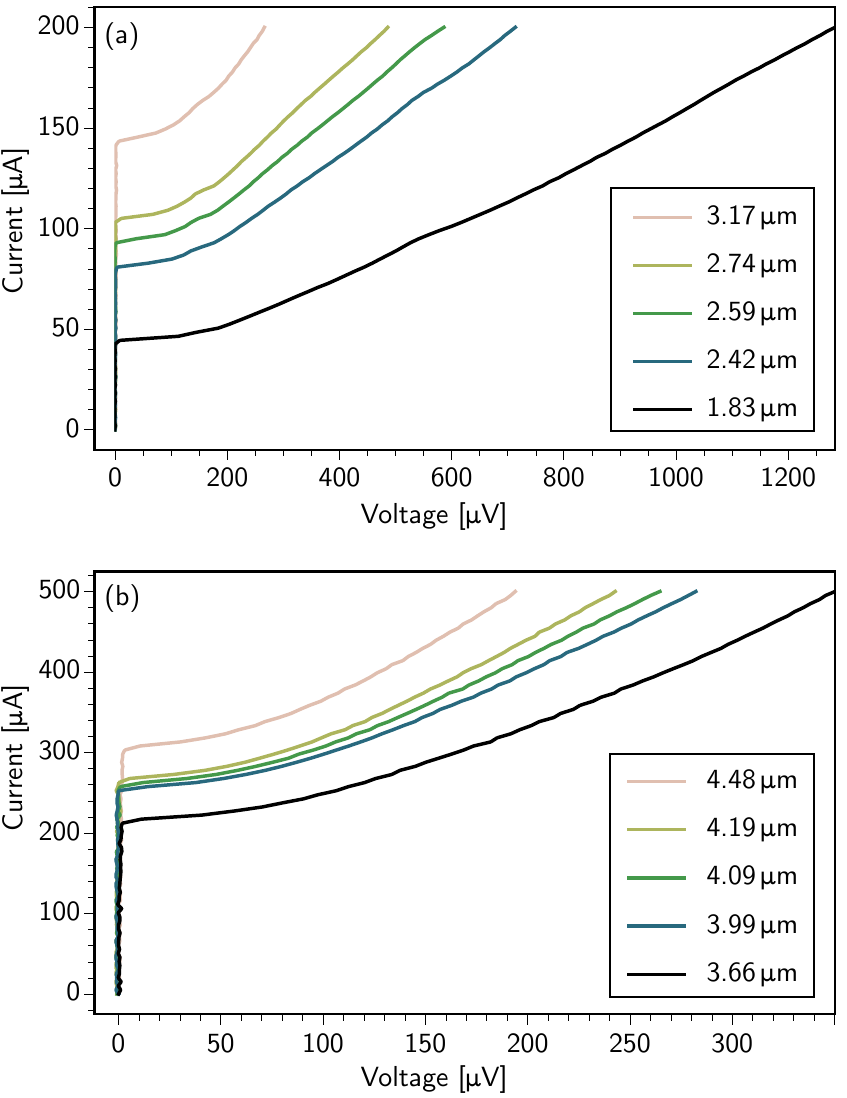}
\caption{JJ IV curves. The SI circuit block used 2.59\,\textmu m junctions with $I_\mathrm{c} \approx 95$\,\textmu A (a), while the DR loop used 4.09\,\textmu m junctions with $I_\mathrm{c} \approx 263$\,\textmu A (b).}
\label{fig:jj_ivs}
\end{figure}
Sample JJ IV curves are presented in Fig.\,\ref{fig:jj_ivs}. The JJs used in the synaptic block all had diameters of 2.59\,\textmu m [Fig.\,\ref{fig:jj_ivs} (a)], while the JJs used in the DR block had diameters of 4.09\,\textmu m [Fig.\,\ref{fig:jj_ivs} (b)], corresponding to critical currents of about 95\,\textmu A and 263\,\textmu A respectively. This varied considerably across the wafer, as evidenced by the range in biases used for the synapses in Fig.\,\ref{fig:big_table}. The critical current density, $J_\mathrm{c}$, of the fab process was measured to be approximately 1.7\,kA/cm$^2$.

\begin{figure*}[t!]
\centering
\includegraphics[width=17.2cm]{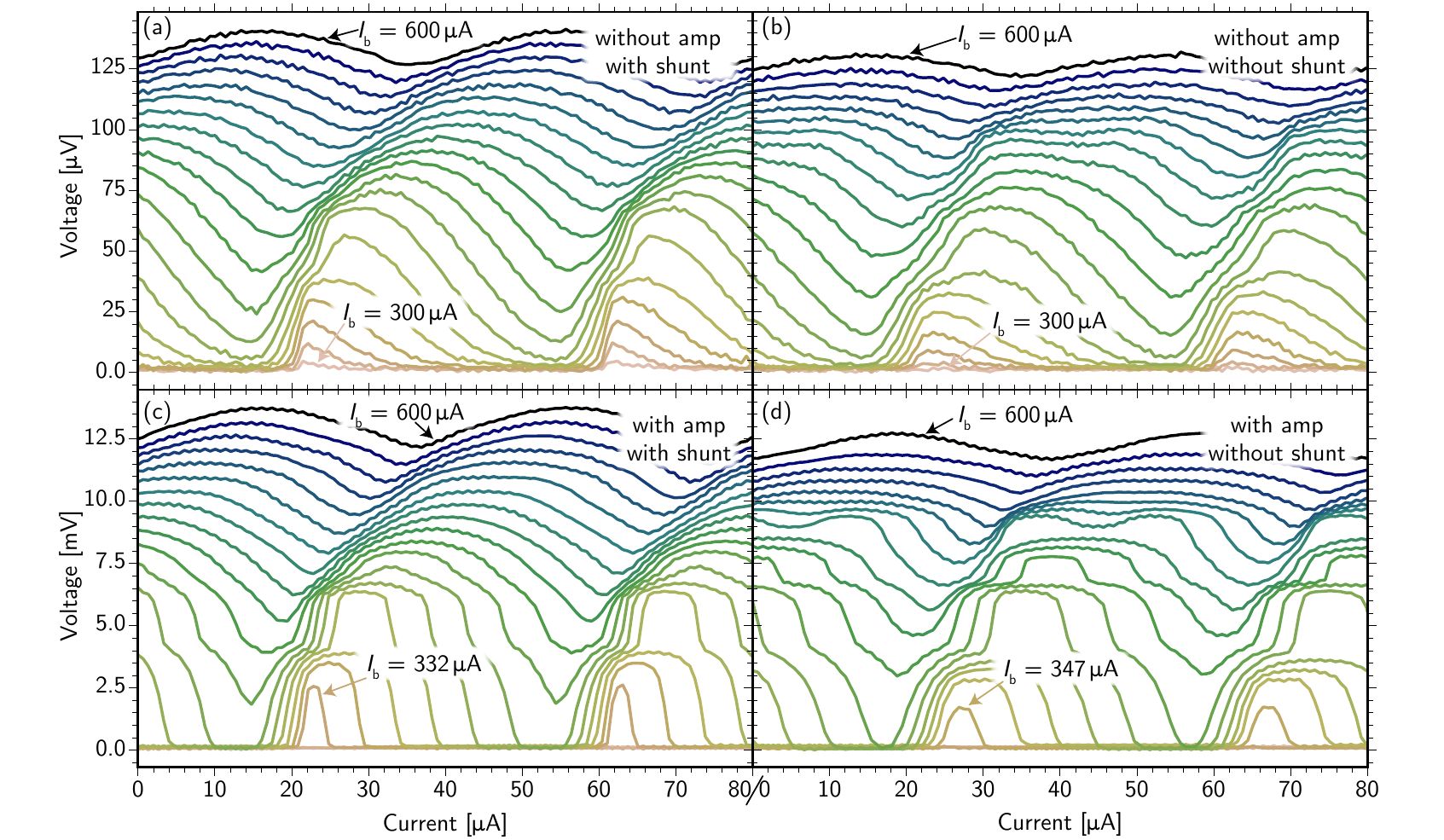}
\caption{Stand-alone (diagnostic) SQUID response as a function of current in what would be the SI loop. The SQUIDs used in the DR loops for the measured synapses correspond to the case shown in (d), but (a) is the device that will be used in for future systems.}
\label{fig:sq_incoil}
\end{figure*}
The response of stand-alone SQUIDs (essentially the DR loop without the synaptic block attached) are shown in Fig.\,\ref{fig:sq_incoil} under several measurement conditions. The $x$-axis is the current sent through a wire inductively coupled to the SQUID loop, mimicking the current in the SI loop of a synapse. Figure\,\ref{fig:sq_incoil}(a), which lacks an amplifier and has a resistive shunt on the SQUID to limit $LC$ resonance effects, shows the results closest to the expected output of an asymmetric SQUID. The synaptic circuits were unfortunately saddled with the worst scenario [Fig.\,\ref{fig:sq_incoil}(d)]: no shunt resistor and an amplifier to resolve small signals. We see that a variety of undesirable abrupt features were introduced in this case, which we think are related to noise coupled in by the amplifier driving $LC$ resonances, where the inductance is due to the SQUID washer and input coil of the transformer, and the capacitance is between the washer and the input coil. A synapse-dendrite combination embedded in a neuron would have a response most similar to Fig.\,\ref{fig:sq_incoil}(a), with the addition of a shunt resistor and without the need for a readout amplifier. 

\begin{figure*}[t!]
\centering
\includegraphics[width=17.2cm]{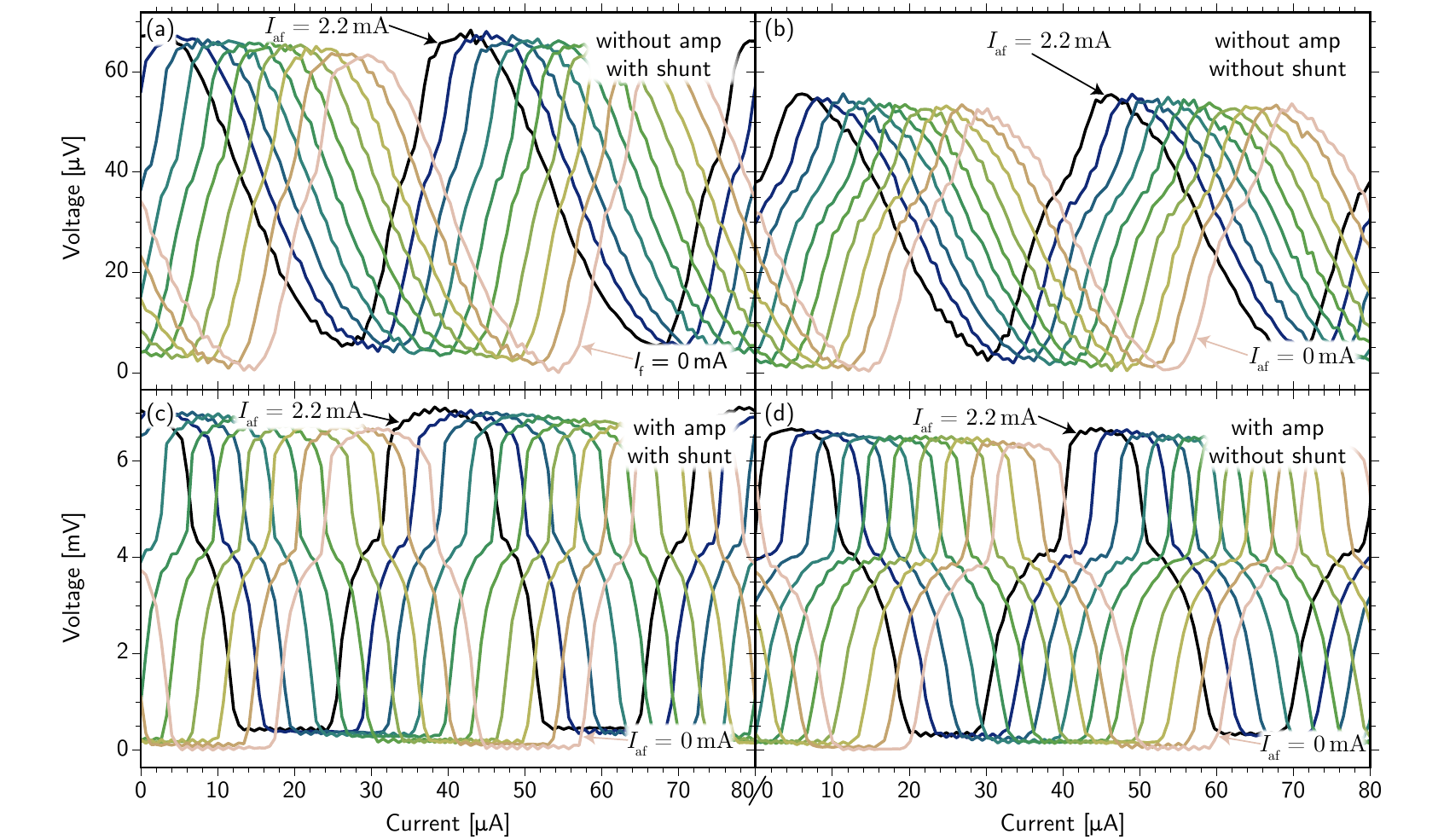}
\caption{SQUID response at a fixed bias current for different values of $I_\mathrm{af}$, demonstrating the ability to tune the phase of the SQUID response.}
\label{fig:sq_af}
\end{figure*}

Figure \ref{fig:sq_af} shows how the SQUID response could be shifted with a second tuning coil carrying $I_\mathrm{af}$. This ability was used throughout data collection to maximize signal amplitude by cancelling the effects background flux in the loop. The addflux coil was also essential for estimating the SI current from $V_\mathrm{sq}$, as shown in the inset of Fig.\,\ref{fig:single_synapse_detail}(a). $I_\mathrm{af}$ was increased from other measurements on this synapse (see entries for the 6.25\,\textmu s, 2.5\,\textmu H synapse in Fig.\,\ref{fig:big_table}) to ensure an approximately linear transfer function between $I_\mathrm{si}$ and $V_\mathrm{sq}$. The transfer function of a DR Loop coupled to a given synapse was measured by fixing the synaptic bias currents and sweeping the addflux bias. Measurements of the period of the SQUID response provided experimental values for mutual inductance between the DR loop and both the SI loop and the addflux coil. Together, these measurements permit the extraction of the SI current from $V_\mathrm{sq}$.

\clearpage
\section{\label{apx:additional_data}Additional Data from Synapses}
\begin{figure}[t!]
\centering
\includegraphics[width=8.6cm]{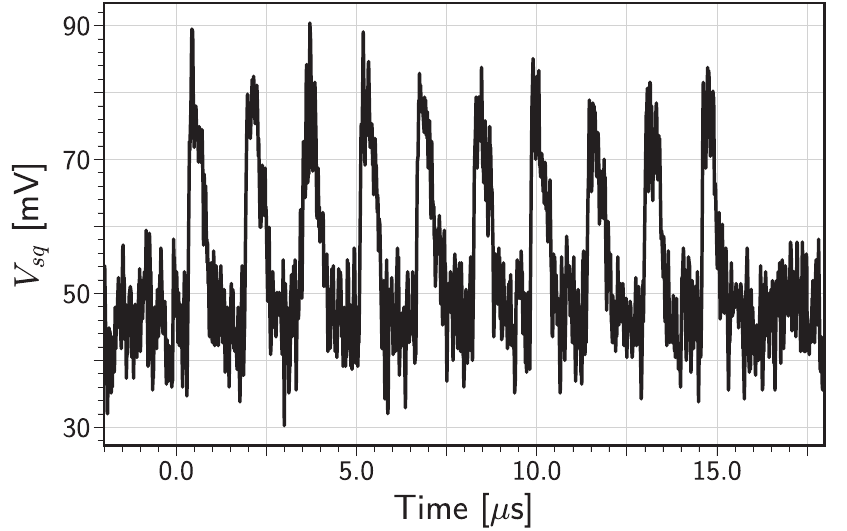}
\caption{Example unaveraged time trace from the 250\,ns synapse.}
\label{fig:unaveraged_time_trace}
\end{figure}

Here we show additional data from various synapses that may be of interest to readers with high stamina. Figure \ref{fig:unaveraged_time_trace} shows a single time trace taken from a synapse wherein no averaging was performed. 

\begin{figure*}
\centering
\includegraphics[width=17.2cm]{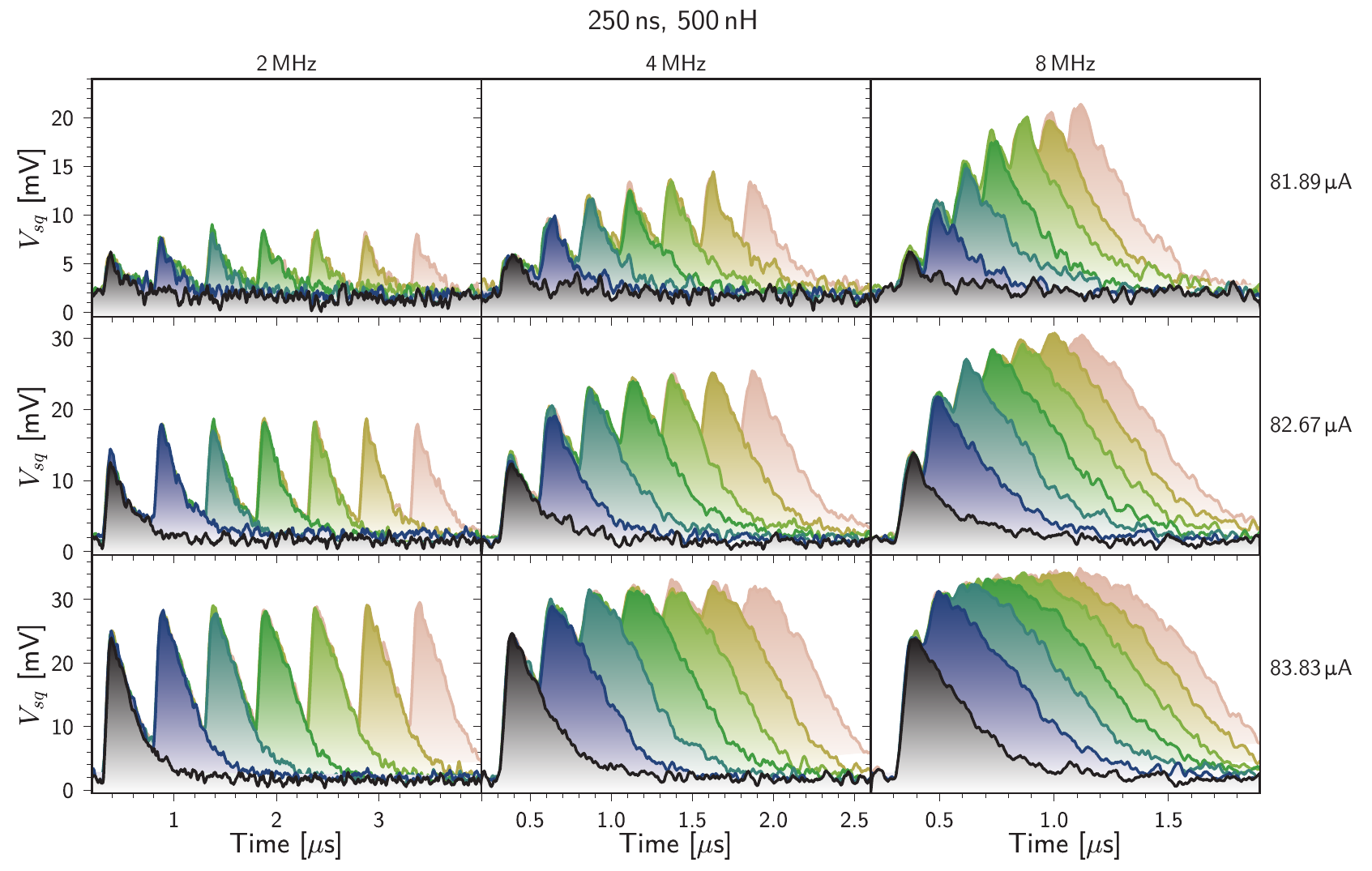}
\caption{250\,ns synapse integrating behavior with different frequency pulse trains (columns), different values of $I_\mathrm{sy}$ (rows), and different numbers of pulses (colors).}
\label{fig:add_data_250ns_500nH_numburst}
\end{figure*}

Figure \ref{fig:add_data_250ns_500nH_numburst} shows time traces from the 250\,ns synapse with bursts of one through seven photonic pulses incident. The frequency of the input pulse train varies across the columns from 2\,MHz to 8\,MHz showing how integration begins to occur above the $L/r$ cutoff frequency. The three rows show the behavior with three different synaptic bias currents applied.

\begin{figure*}[t!]
\centering
\includegraphics[width=17.2cm]{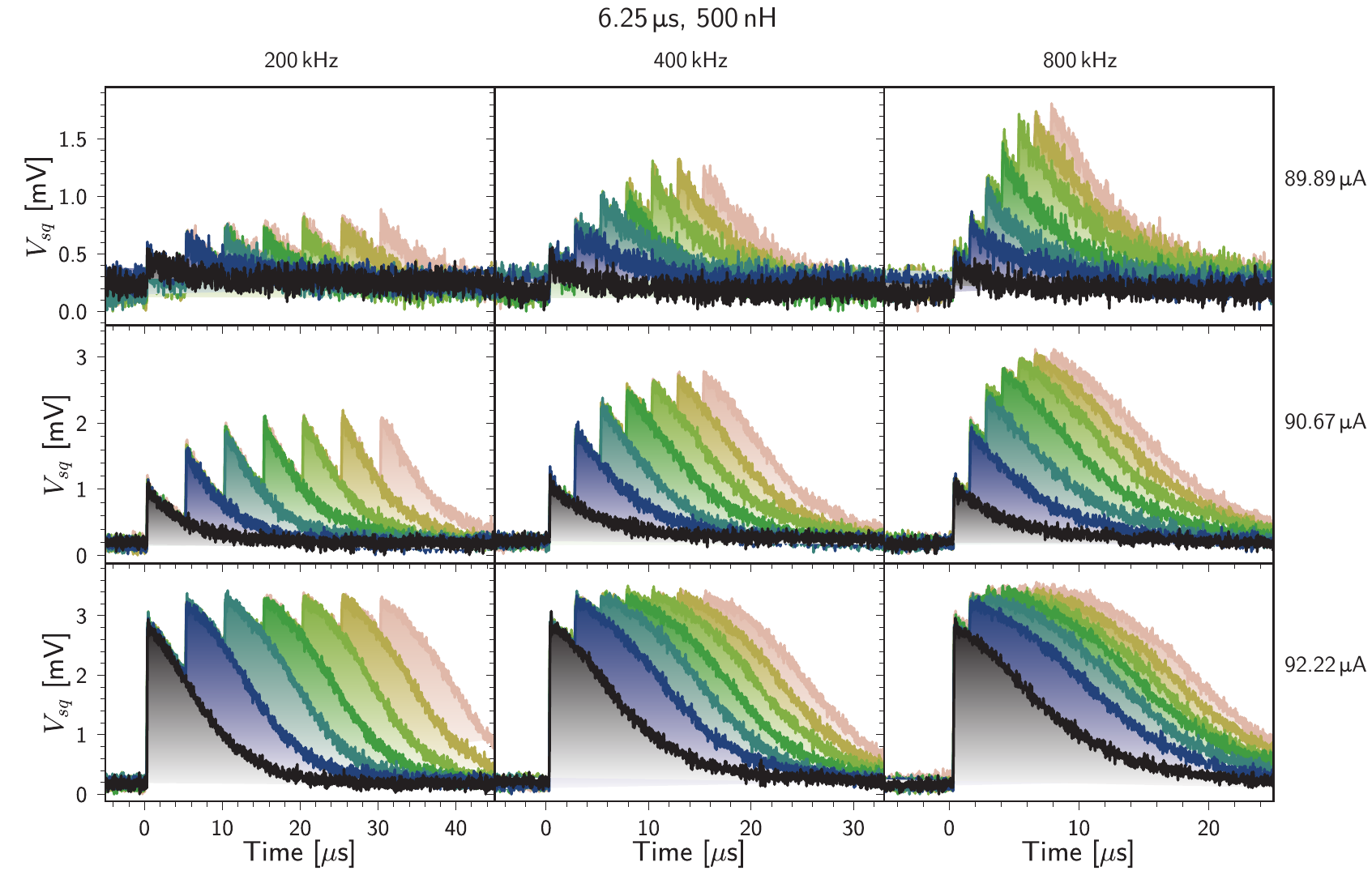}
\caption{6.25\,\textmu s, 500\,nH synapse integrating behavior with different frequency pulse trains (columns), different values of $I_\mathrm{sy}$ (rows), and different numbers of pulses (colors).}
\label{fig:add_data_6p25us_500nH_numburst}
\end{figure*}

Figure \ref{fig:add_data_6p25us_500nH_numburst} shows data similar to Fig.\,\ref{fig:add_data_250ns_500nH_numburst}, except for the 6.25\,\textmu s, 500\,nH synapse. The frequencies have been shifted lower by an order of magnitude to demonstrate the same temporal integration crossover effect.

\begin{figure*}
\centering
\includegraphics[width=17.2cm]{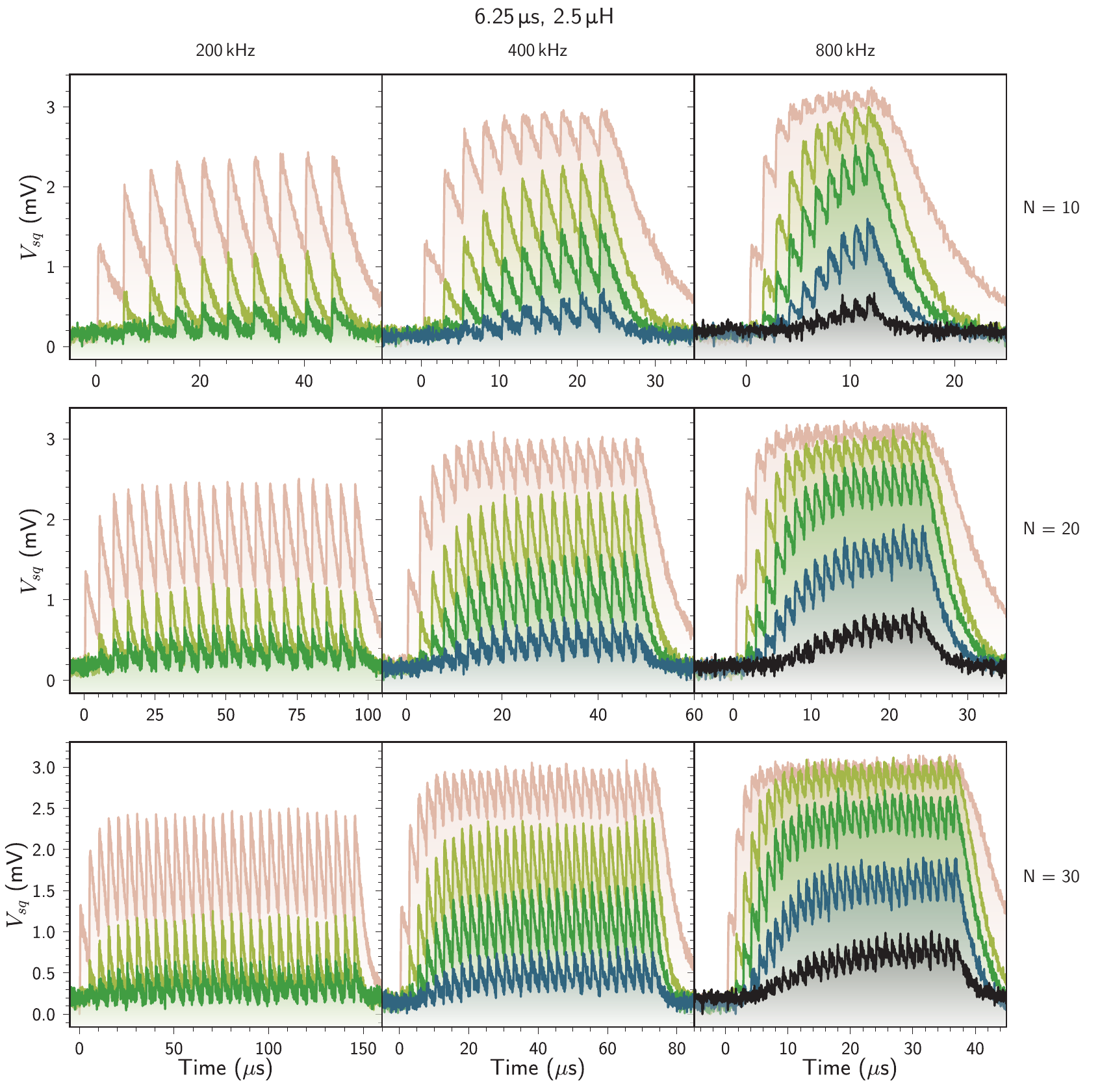}
\caption{6.25\,\textmu s, 2.5\,\textmu H synapse integrating behavior with different frequency pulse trains (columns), different numbers of pulses in a burst (rows), and different synaptic weights (colors).}
\label{fig:add_data_6p25us_2p5uH_longburst}
\end{figure*}

\begin{figure*}
\centering
\includegraphics[width=8.6cm]{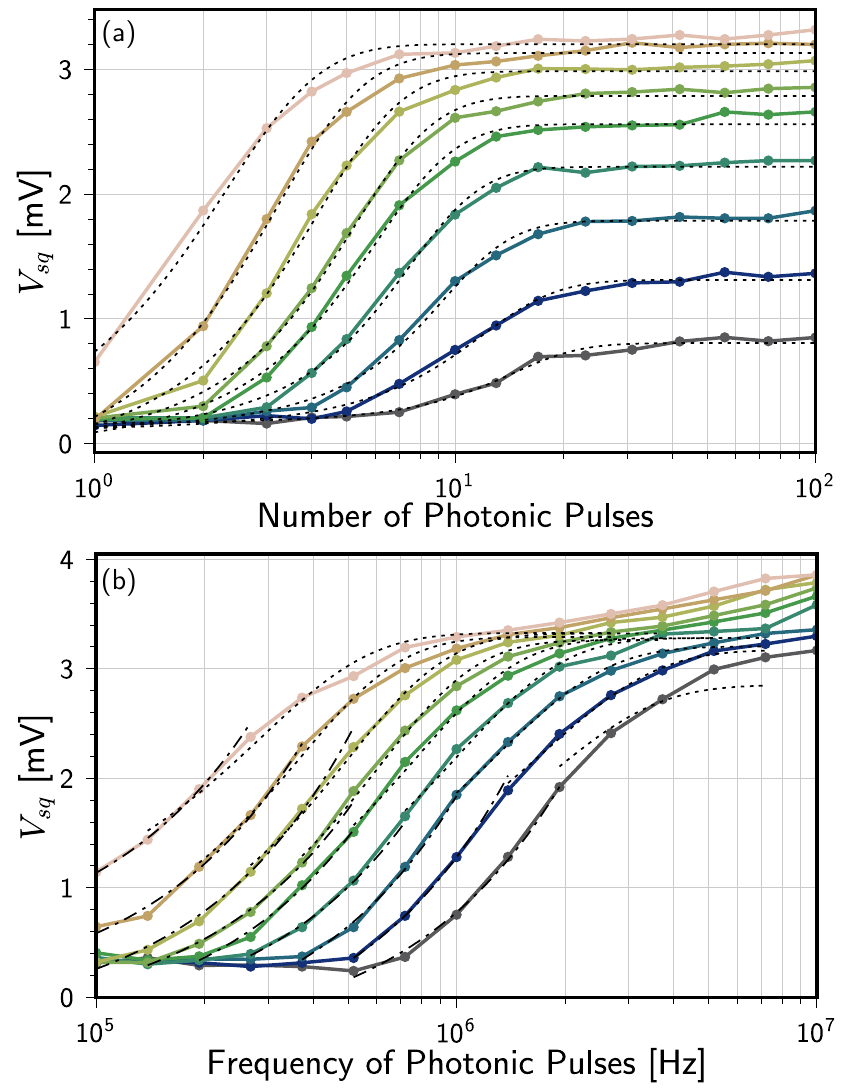}
\caption{Fits to the transfer functions of the 6.25\,ns, 2.5\,\textmu H synapse. (a) Sigmoidal fits to the burst transfer functions for the same cases as Fig.\,\ref{fig:single_synapse_detail}(e). (b) Fits to the frequency transfer functions for the same cases as Fig.\,\ref{fig:single_synapse_detail}(f). In (b) the dashed-dotted lines are linear fits that capture the behavior well for the initial turn-on segment, and the dotted lines are sigmoidal fits that capture the behavior as the curves roll over.}
\label{fig:transfer_function_fits}
\end{figure*}

Figures \ref{fig:add_data_6p25us_2p5uH_longburst} and \ref{fig:transfer_function_fits} investigate the 6.25\,\textmu s, 2.5\,\textmu H synapse. Figure \ref{fig:add_data_6p25us_2p5uH_longburst} shows integrating behavior. The three columns show three frequencies while the three rows show different numbers of pulses in the incident photonic pulse train. Within each panel, several values of synaptic weight are shown. As the pulse train is shifted to higher frequency, weaker synaptic weights are sufficient to bring the integrated signal above the noise.

Figure \ref{fig:transfer_function_fits} shows the transfer functions measured from this synapse, just as in Fig.\,\ref{fig:single_synapse_detail}(e) and (f), but here we have included fits to the data. In Fig.\,\ref{fig:transfer_function_fits}(a) fits to a sigmoidal functional form are shown as dotted lines. The sigmoid functional form is given by
\begin{equation}
    \label{eq:sigmoid}
    V_\mathrm{sq}(N_\mathrm{ph}) = \left(A - A_0 \right) \left[ 1 - \left( e^{ (N_\mathrm{ph} - N_0) / w} + 1 \right)^{-1} \right] + A_0,
\end{equation}
where $A$, $A_0$, $N_0$ and $w$ have been fit using the Scipy {\fontfamily{cmtt}\selectfont curve\_fit} function. Figure \ref{fig:transfer_function_fits}(b) shows the transfer functions versus the frequency of photonic pulses. In this case, the sigmoid form of Eq.\,\ref{eq:sigmoid} provides a reasonable fit of the roll-over section of the curves (dotted lines), while the initial turn-on section is better fit with a linear function. Those fits are shown with dashed dotted lines.

\end{document}